\documentclass[aps, prc, twocolumn, superscriptaddress]{revtex4-1}

\usepackage{graphicx}
\usepackage{longtable}
\usepackage{subfigure}
\usepackage{amsmath}

\begin{document}

\title{Decay-assisted collinear resonance ionization spectroscopy: \\Application to neutron-deficient francium}

\author{K. M. Lynch}
\email{kara.marie.lynch@cern.ch}
\affiliation{School of Physics and Astronomy, The University of Manchester, Manchester, M13 9PL, United Kingdom}
\affiliation{ISOLDE, PH Department, CERN, CH-1211 Geneva-23, Switzerland}
\affiliation{Instituut voor Kern- en Stralingsfysica, KU Leuven, B-3001 Leuven, Belgium}

\author{J. Billowes}
\affiliation{School of Physics and Astronomy, The University of Manchester, Manchester, M13 9PL, United Kingdom}

\author{M. L. Bissell}
\affiliation{Instituut voor Kern- en Stralingsfysica, KU Leuven, B-3001 Leuven, Belgium}

\author{I. Budin\u{c}evi\'{c}}
\affiliation{Instituut voor Kern- en Stralingsfysica, KU Leuven, B-3001 Leuven, Belgium}

\author{T.E. Cocolios}
\affiliation{School of Physics and Astronomy, The University of Manchester, Manchester, M13 9PL, United Kingdom}
\affiliation{ISOLDE, PH Department, CERN, CH-1211 Geneva-23, Switzerland}

\author{R.P. De Groote}
\affiliation{Instituut voor Kern- en Stralingsfysica, KU Leuven, B-3001 Leuven, Belgium}

\author{S. De Schepper}
\affiliation{Instituut voor Kern- en Stralingsfysica, KU Leuven, B-3001 Leuven, Belgium}

\author{V.N. Fedosseev}
\affiliation{EN Department, CERN, CH-1211 Geneva 23, Switzerland}

\author{K.T. Flanagan}
\affiliation{School of Physics and Astronomy, The University of Manchester, Manchester, M13 9PL, United Kingdom}

\author{S. Franchoo}
\affiliation{Institut de Physique Nucl\'{e}aire d'Orsay, F-91406 Orsay, France}

\author{R.F. Garcia Ruiz}
\affiliation{Instituut voor Kern- en Stralingsfysica, KU Leuven, B-3001 Leuven, Belgium}

\author{H. Heylen}
\affiliation{Instituut voor Kern- en Stralingsfysica, KU Leuven, B-3001 Leuven, Belgium}

\author{B.A. Marsh}
\affiliation{EN Department, CERN, CH-1211 Geneva 23, Switzerland}

\author{G. Neyens}
\affiliation{Instituut voor Kern- en Stralingsfysica, KU Leuven, B-3001 Leuven, Belgium}

\author{T.J. Procter}
\email{Present address: TRIUMF, Vancouver, British Columbia, V6T 2A3, Canada}
\affiliation{School of Physics and Astronomy, The University of Manchester, Manchester, M13 9PL, United Kingdom}

\author{R.E. Rossel}
\affiliation{EN Department, CERN, CH-1211 Geneva 23, Switzerland}
\affiliation{Institut f\"{u}r Physik, Johannes Gutenberg-Universit\"{a}t Mainz, D-55128 Mainz, Germany}

\author{S. Rothe}
\affiliation{EN Department, CERN, CH-1211 Geneva 23, Switzerland}
\affiliation{Institut f\"{u}r Physik, Johannes Gutenberg-Universit\"{a}t Mainz, D-55128 Mainz, Germany}

\author{I. Strashnov}
\affiliation{School of Physics and Astronomy, The University of Manchester, Manchester, M13 9PL, United Kingdom}

\author{H.H. Stroke}
\affiliation{Department of Physics, New York University, NY, New York 10003, USA}

\author{K.D.A. Wendt}
\affiliation{Institut f\"{u}r Physik, Johannes Gutenberg-Universit\"{a}t Mainz, D-55128 Mainz, Germany}

\date{\today}

\begin{abstract}
This paper reports on the hyperfine-structure and radioactive-decay studies of the neutron-deficient francium isotopes $^{202-206}$Fr performed with the Collinear Resonance Ionization Spectroscopy (CRIS) experiment at the ISOLDE facility, CERN. The high resolution innate to collinear laser spectroscopy is combined with the high efficiency of ion detection to provide a highly-sensitive technique to probe the hyperfine structure of exotic isotopes. The technique of decay-assisted laser spectroscopy is presented, whereby the isomeric ion beam is deflected to a decay spectroscopy station for alpha-decay tagging of the hyperfine components. Here, we present the first hyperfine-structure measurements of the neutron-deficient francium isotopes $^{202-206}$Fr, in addition to the identification of the low-lying states of $^{202,204}$Fr performed at the CRIS experiment.
\end{abstract}

\pacs{}

\maketitle

\section{Introduction}

Recent advances in high-resolution laser spectroscopy have resulted in the ability to measure short-lived isotopes with yields of less than 100 atoms per second~\cite{Cheal2010a, Blaum2013}. The Collinear Resonance Ionization Spectroscopy (CRIS) experiment~\cite{Procter2013}, located at the ISOLDE facility, CERN, aims to push the limits of laser spectroscopy further, performing hyperfine-structure measurements on isotopes at the edges of the nuclear landscape. It provides a combination of high-detection efficiency, high resolution and ultra-low background, allowing measurements to be performed on isotopes with yields below, in principle, one atom per second. 

The first optical measurements of francium were performed in 1978. Liberman identified the 7s~$^2$S$_{1/2} \rightarrow$ 7p~$^2$P$_{3/2}$ atomic transition, performing hyperfine-structure and isotope-shift measurements first with low-resolution~\cite{Liberman1978} and later with high-resolution laser spectroscopy~\cite{Liberman1980}. The wavelength of this transition $\lambda$(D2) = 717.97(1)~nm was in excellent agreement with the prediction of Yagoda~\cite{Yagoda1932}, made in 1932 before francium was discovered. Further measurements of francium followed in the next decade. High-resolution optical measurements were performed on both the 7s~$^2$S$_{1/2} \rightarrow$ 7p~$^2$P$_{3/2}$ atomic transition~\cite{Coc1985, Touchard1984, Bauche1986}, as well as the 7s~$^2$S$_{1/2} \rightarrow$ 8p~$^2$P$_{3/2}$ transition~\cite{Duong1987}, along with transitions into high-lying Rydberg states~\cite{Andreev1987}. The CRIS technique, a combination of collinear laser spectroscopy and resonance ionization was originally proposed by Kudriavtsev in 1982~\cite{Kudriavtsev1982}, but the only experimental realization of the technique was not performed until 1991 on ytterbium atoms~\cite{Schulz1991}.

The ability to study the neutron-deficient francium (Z~=~87) isotopes at the CRIS beam line offers the unique opportunity to answer questions arising from the study of the nuclear structure in this region of the nuclear chart. As the isotopes above the Z~=~82 shell closure become more neutron deficient, a decrease in the excitation energy of the ($\pi $1i$_{13/2}$)$_{13/2^+}$, ($\pi$2f$_{7/2}$)$_{7/2^+}$, ($\nu$1i$_{13/2}$)$_{13/2^+}$ and ($\pi$3s$^{-1}_{1/2}$)$_{1/2^+}$ states is observed. In $^{185}$Bi (Z~=~83) and $^{195}$At (Z~=~85), the ($\pi$3s$^{-1}_{1/2}$)$_{1/2^+}$ deformed intruder state has been observed to be the ground state~\cite{Davids1996, Kettunen2003}.

Recent radioactive-decay measurements suggest the existence of a ($\pi$3s$^{-1}_{1/2}$)$_{1/2^+}$ proton intruder state for $^{203}$Fr and, with a lower excitation energy, for $^{201}$Fr, suggesting that this state may become the ground state in $^{199}$Fr~\cite{Uusitalo2005, Jakobsson2012}. The intruder configurations polarize the nucleus, creating significant deformation. From the study of the nuclear structure of the neutron-deficient francium isotopes towards $^{199}$Fr (by measuring the magnetic dipole moments and change in mean-square charge radii of the ground and isomeric states), the quantum configuration of the states and the shape of the nuclei can be investigated.

Radioactive-decay measurements on the neutron-deficient francium isotopes have aimed to determine the level structure of the low-lying nuclear states, but their exact nature is still unknown~\cite{Huyse1992, Uusitalo2005, Jakobsson2012, Jakobsson2013}. High-resolution collinear laser spectroscopy has allowed determination of the ground-state properties of $^{204,205,206}$Fr~\cite{Voss2013}, confirming the tentative spin assignments. The spin of $^{205}$Fr was measured to be 9/2$^-$, the ground-state spins of $^{204,206}$Fr were confirmed as 3$^{(+)}$, but the low-lying spin (7$^+$) and (10$^-$) isomers are still under investigation. 

General methods of isomer identification have already been achieved with in-source laser spectroscopy~\cite{Fedosseev2012a} (and references therein). In the case of $^{68,70}$Cu~\cite{Weissmann2002}, following the selection of isomeric beams, experiments such as Coulomb excitation~\cite{Stefanescu2007} and mass measurements~\cite{VanRoosbroeck2004} have been performed. However, these experiments suffered from isobaric contamination, as well as significant ground-state contamination due to the Doppler broadening of the hyperfine resonances of each isomer~\cite{Cheal2010a}. One way of addressing the difficulties of in-source laser spectroscopy (isobaric contamination, Doppler broadening, pressure broadening) is selecting the ground or isomeric state of interest by resonance ionization in a collinear geometry. 

In a sub-Doppler geometry, the process of isomer selective resonance laser ionization~\cite{Fedosseev2012a} can result in a high-purity isomeric beam. Deflection of the pure-state ion beam to the decay spectroscopy station allows identification of the hyperfine component with alpha-decay spectroscopy.

\section{Experimental technique}

\begin{figure*}
\includegraphics[width = \textwidth]{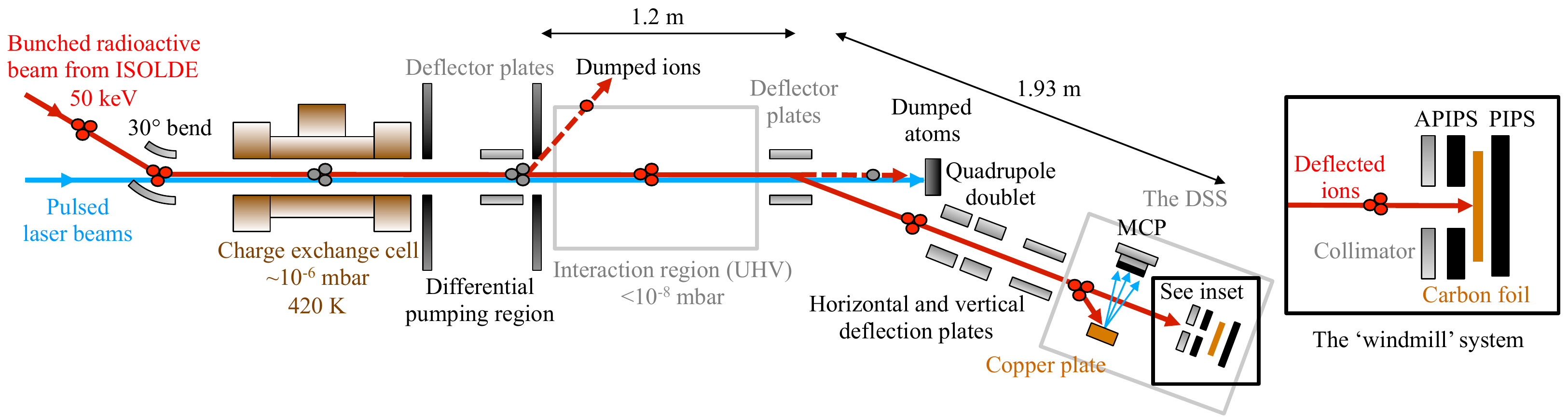}
\caption{\label{fig:CRIS_beamline}Schematic diagram of the CRIS beam line. Laser ions can be deflected to a copper plate and the corresponding secondary electrons detected by the MCP, or implanted into a carbon foil for alpha-decay spectroscopy. (Inset) The decay spectroscopy station (DSS) `windmill' system for alpha-decay tagging.}
\end{figure*}

Radioactive ion beams of francium were produced at the ISOLDE facility, CERN~\cite{Kugler2000} by impinging 1.4~GeV protons onto a thick UC$_x$ target (up to 2~$\mu$A integrated proton current). The radioisotopes were surface ionized through interaction with the rhenium coating on the hot (2400~K) tantalum transfer line and extracted from the target-ion source at 50~keV. The isotope of interest was mass selected using the high-resolution HRS separator and bunched (at 31.25~Hz) with the radio-frequency cooler-buncher ISCOOL~\cite{Jokinen2003, Mane2009}. The bunched-ion beam was deflected into the CRIS beam line and transported through a potassium-vapour charge exchange cell (CEC) (420~K, $\sim$10$^{-6}$~mbar chamber pressure, 6$\times 10^{-4}$~mbar vapour pressure~\cite{Handbook2013}) to be neutralized. In the 1.2~m long interaction region, the arrival of the atomic bunch was synchronized with two co-propagating pulsed laser beams to excite the state of interest followed by ionization in a step-wise scheme. The temporal length of the atomic bunch was 2-3~$\mu$s, corresponding to a spatial length of 45-70~cm. To reduce the background signal resulting from non-resonant collisional ionization, the interaction region aims at ultra-high vacuum (UHV) conditions. A pressure of $<$10$^{-8}$~mbar was achieved during this experiment. A schematic diagram of the CRIS beam line is shown in Fig.~\ref{fig:CRIS_beamline}.

\subsection{Collinear resonance ionization spectroscopy}

The resonant excitation step from the 7s~$^2$S$_{1/2}$ electronic ground state to the 8p~$^2$P$_{3/2}$ state was probed with 422.7-nm light. The laser light of this resonant step was provided by a narrow-band titanium:sapphire (Ti:Sa) laser of the ISOLDE RILIS installation~\cite{Fedosseev2012b,Rothe2013}, pumped by the second harmonic output of a Nd:YAG laser (Model: Photonics Industries DM-60-532, 10~kHz). The fundamental output from the tuneable Ti:Sa laser was frequency doubled using a BBO crystal to produce the required 422.7-nm laser light. The light was fibre-coupled into the CRIS beam line through 35~m of multimode optical fibre ($\sim$100~mW output). The laser linewidth of 1.5~GHz limited the resolution achieved in the present experiment, allowing only the lower-state (7s~$^2$S$_{1/2}$) splitting to be fully resolved. The second (non-resonant) transition from the 8p~$^2$P$_{3/2}$ state to the continuum was driven using 1064-nm light. This light was produced by a fundamental Nd:YAG laser (Model: Quanta-Ray LAB 130, operated at 31.25~Hz) next to the CRIS setup, temporally overlapped with the 422.7-nm laser beam and aligned through the laser/atom interaction region. The standard repetition rate of the RILIS lasers (10~kHz) limited the repetition rate of the 1064-nm laser light to 31.25~Hz (one out of every 320 pulses of 422.7-nm laser light was utilized). The bunching of the ion beam with ISCOOL was matched to the lower repetition rate of 31.25~Hz to overlap the atom bunch with the two laser pulses every 32~ms. 

The synchronization of the first- and second-step laser pulses and the release of the ion bunch from ISCOOL was controlled by a Quantum Composers digital delay generator (Model: QC9258). The 10~kHz pulse generator of the Ti:Sa pump laser acted as the master clock, triggering the delay generator to output a sequence of TTL pulses to synchronize the 1064-nm laser light and the ion bunch with the 422.7-nm light, allowing resonance ionization of the francium atoms to occur. The laser ions were detected by a micro-channel plate (MCP) housed in the decay spectroscopy station (DSS). The electronic signal from the MCP was digitized by a LeCroy oscilloscope (Model: WavePro 725Zi, 2~GHz bandwidth, 8~bit ADC, 20~GS/s), triggered by the digital delay generator. The data were transferred from the oscilloscope using a LabVIEW\texttrademark~program.

The frequency of the resonant excitation step, the 422.7-nm laser light, was scanned to study the 7s~$^2$S$_{1/2} \rightarrow$ 8p~$^2$P$_{3/2}$ atomic transition. The scanning and stabilization of the frequency was controlled by the RILIS Equipment Acquisition and Control Tool (REACT), a LabVIEW control program package that allows for remote control, equipment monitoring and data acquisition~\cite{Rossel2013}. This was achieved by controlling the etalon tilt angle inside the Ti:Sa laser resonator to adjust the laser wavelength, which was measured with a HighFinesse wavemeter (Model: WS7), calibrated with a frequency stabilised HeNe laser. The francium experimental campaign at CRIS marked the first implementation of the REACT framework for external users. The remote control LabVIEW interface for the Ti:Sa laser ran locally at the CRIS setup, allowing independent laser scanning and control.

\subsection{Decay-assisted laser spectroscopy}

The technique of decay-assisted collinear laser spectroscopy was further developed at the CRIS beam line to take advantage of the ultra-pure ion beams produced by resonance ionization in a collinear geometry~\cite{Lynch2013}. The selectivity from resonance ionization of an isotope is a result of the selectivity of the Lorentzian profile of the natural linewidth ($\sim$12.5~MHz) of the state and the Gaussian profile of the laser linewidth ($\sim$1.5~GHz). At a frequency separation of 4~GHz, the Gaussian component falls to 1\% of its peak intensity and the selectivity is dominated by the natural linewidth of the state. Thus, the maximum selectivity from resonance ionization is given by Eq.~(\ref{eq:selectivity}),
\begin{equation} S = \prod_{n=1}^N \Big( \frac{\Delta \omega_{\textnormal{AB},n}}{\Gamma_n} \Big)^2 = \prod_{n=1}^N S_n ,\label{eq:selectivity} \end{equation}
where $\Delta \omega_{\textnormal{AB}}$ is the separation in frequency of the two states (A and B), $\Gamma_n$ is the FWHM of the natural linewidth of the state, $S_n$ is the selectivity of the transition and $N$ is the number of transitions used. The total selectivity of a resonance ionization process is given by the product of the individual selectivities. In the case of the two states being the ground state and isomer, the selectivity can be calculated from Eq.~(\ref{eq:selectivity}). When the two states are the isotope of interest and contamination from a neighbouring isotope, additional selectivity can be gained from the kinematic shift since the laser is overlapped with an accelerated beam.

In addition to hyperfine-structure studies with ion detection, the decay spectroscopy station can be used to identify the hyperfine components of overlapping structures. This allows the hyperfine structure of two states to be separated by exploiting their characteristic radioactive-decay mechanisms. This results in a smaller error associated with the hyperfine parameters, and a better determination of the extracted nuclear observables.

The decay spectroscopy station (DSS) consists of a rotatable wheel implantation system~\cite{Rajabali2013}. It is based on the design from KU Leuven~\cite{DendoovenThesis} (Fig.~1 of Ref.~\cite{Andreyev2010}), which has provided results in a number of successful experiments~\cite{Elseviers2013} (and references therein). The wheel holds 9 carbon foils, produced at the GSI target laboratory~\cite{Lommel2002}, with a thickness of 20(1)~$\mu$g~cm$^{-2}$~($\sim$90~nm) into which the ion beam is implanted (at a depth of $\sim$25~nm). 

Two Canberra Passivated Implanted Planar Silicon (PIPS) detectors for charged-particle detection (e.g.~alpha, electron, fission fragments) are situated on either side of the implantation carbon foil, as shown in Fig.~\ref{fig:CRIS_beamline}. One PIPS detector (Model: BKA 300-17 AM, thickness 300~$\mu$m) sits behind the carbon foil and another annular PIPS (APIPS) (Model: BKANPD 300-18 RM, thickness 300~$\mu$m, with an aperture of 4~mm) is placed in front of the carbon foil. The detectors are connected to charge sensitive Canberra preamplifiers (Model: 2003BT) via a UHV type-C sub-miniature electrical feed-through. 

Laser-produced ions from the interaction region in the CRIS beam line are deflected to the DSS by applying a potential difference between a pair of vertical electrostatic plates, see Fig.~\ref{fig:CRIS_beamline}. The deflected ion beam is implanted into the carbon foil, after passing though a collimator with a 4~mm aperture and the APIPS detector. The collimator shields the APIPS detector from direct implantation of radioactive ions into the silicon wafer, see Fig.~\ref{fig:CRIS_beamline}. Decay products from the carbon foil can be measured by either the APIPS or PIPS detector, with a total solid-angle coverage of 63\% (simulated, assuming a uniform distribution of implanted activity). Operation of the single APIPS detector during the experiment gave an alpha-detection efficiency of 25\%. An electrical contact is made to the collimator, allowing the current generated by the ion beam when it strikes the collimator to be measured and the plate to be used as a beam monitoring device. When it is not in use, it is electrically grounded to avoid charge build-up. A Faraday cup is installed in the location of one of the carbon foils. This copper plate (thickness 0.5~mm, diameter 10~mm) is electrically isolated from the steel wheel by PEEK rings and connected by a spot-welded Kapton cable attached to a rotatable BNC connection in the centre of the wheel~\cite{Rajabali2013}.

The alpha-decay spectroscopy data is acquired with a digital data acquisition system (DAQ), consisting of XIA digital gamma finder (DGF) revision D modules~\cite{Hennig2007}. Each module has four input channels with a 40~MHz sampling rate. Signals fed into the digital DAQ are self-triggered with no implementation of master triggers.

Due to the reflective surface of the inside of the vacuum chambers, a significant fraction of 1064-nm laser light was able to scatter into the silicon detectors. Despite the collimator in front of the APIPS detector to protect it from ion implantation (and laser light), the infra-red light caused a shift in the baseline of the signal from the silicon detector. This required the parameters for the DGF modules to be adjusted to account for this effect online, since the reflections were due to the particular setup of the experiment (power and laser-beam path).

The low energy resolution of the APIPS detector was associated with the necessity of optimizing the DGF parameters online with the radioactive $^{221}$Fr (t$_{1/2}$ = 4.9(2)~min). In addition, a fluctuating baseline resulting from the changing power of the 1064-nm laser light meant that only a resolution of 30~keV at 6.341~MeV was achieved. This however was sufficient to identify the characteristic alpha decays of the neutron-deficient francium isotopes under investigation.

\section{Results}

The hyperfine structures of the neutron-deficient francium isotopes $^{202-206}$Fr were measured with collinear resonance ionization spectroscopy, with respect to the reference isotope $^{221}$Fr. This paper follows the recent publication reporting the hyperfine-structure studies of $^{202,203,205}$Fr~\cite{Flanagan2013}. During the experimental campaign, the neutron-rich francium isotopes $^{218m,219,229,231}$Fr were also studied. A detailed description of the nature of these isotopes will be the topic of a future publication~\cite{Budincevic2014}. 

The resonance spectrum of the 7s~$^2$S$_{1/2}$ $\rightarrow$ 8p~$^2$P$_{3/2}$ transition was fit with a $\chi^2$-minimization routine. The hyperfine $A_{P_{3/2}}$ factor was fixed to the ratio of the 7s~$^2$S$_{1/2}$ $\rightarrow$ 8p~$^2$P$_{3/2}$ transition of $A_{P_{3/2}}/A_{S_{1/2}}=+22.4/6209.9$, given in literature~\cite{Duong1987}. For the 8p~$^2$P$_{3/2}$ state, the hyperfine $B_{P_{3/2}}$ factor is small enough to have no impact on the fit to the data, and was consequently set to zero~\cite{Cocolios2013b}. 

\begin{figure}
\includegraphics[width = \columnwidth]{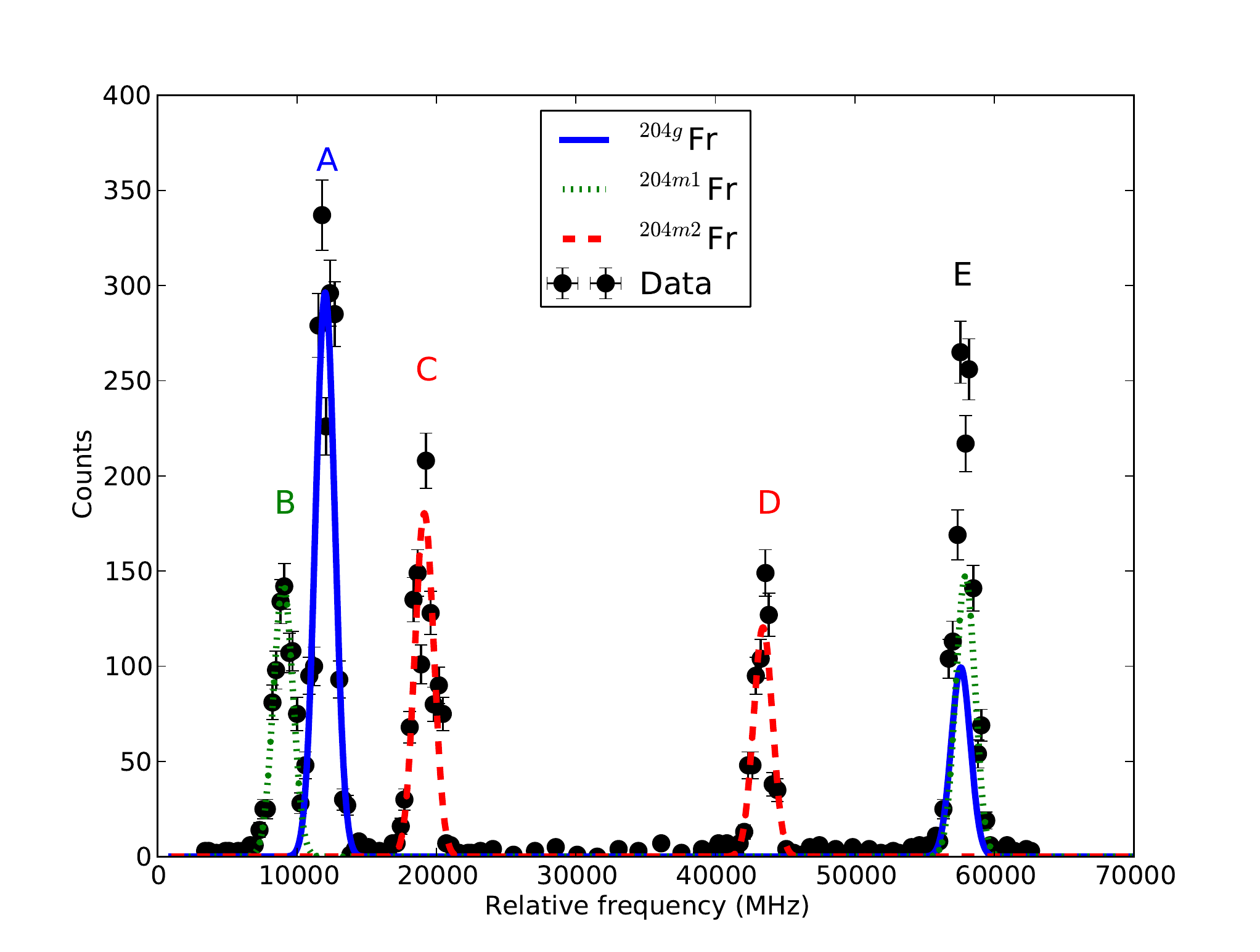}
\caption{\label{fig:CRIS_Fr204} Collinear resonance ionization spectroscopy of $^{204}$Fr relative to $^{221}$Fr. The hyperfine structure of the 3$^{(+)}$ ground state of $^{204g}$Fr is shown in blue, the 7$^+$ state of $^{204m1}$Fr is shown in green and the (10$^-$) state of $^{204m2}$Fr is shown in red.}
\end{figure}

\begin{figure}
\includegraphics[width = \columnwidth]{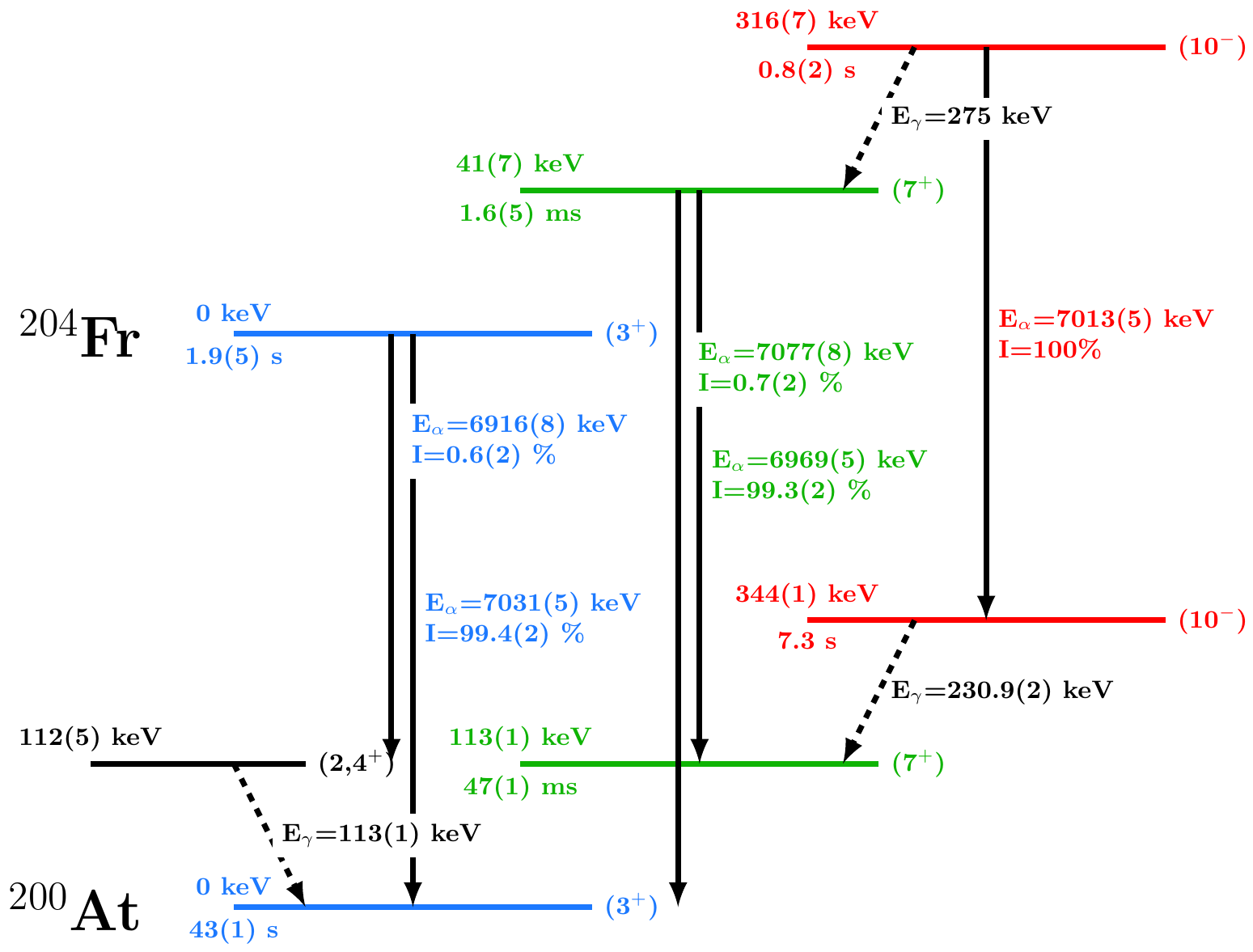}
\caption{\label{fig:Fr204_alpha_decay} The radioactive decay of $^{204}$Fr and its isomers~\cite{Huyse1992, Uusitalo2005, Jakobsson2013}.}
\end{figure}

The intensities of the hyperfine transitions $S_{FF'}$ between hyperfine levels $F$ and $F'$ (with angular momentum $J$ and $J'$ respectively) are related to the intensity of the underlying fine structure transition $S_{JJ'}$~\cite{Blaum2013}. The relative intensities of the hyperfine transitions are given by
\begin{equation}\label{eq:hf_intensities} \frac{S_{FF'}}{S_{JJ'}} = (2F +1) (2F' +1) \begin{Bmatrix}
F & F' & 1 \\
J' & J & I
 \end{Bmatrix} ^2 ,\end{equation}
where $\{ \ldots \}$ denotes the Wigner 6-$j$ coefficient. Although these theoretical intensities are only strictly valid for closed two-level systems, and there was jitter on the temporal overlap of the two laser pulses in the interaction region, they were used as currently the most reliable estimate.

The $A_{S_{1/2}}$ factor and the centroid frequency of the hyperfine structure were determined for each scan individually. For isotopes with multiple scans, a weighted mean for the $A_{S_{1/2}}$ factor and the centroid frequency were calculated based on the error of the fits. The uncertainty attributed to the $A_{S_{1/2}}$ factor was calculated as the weighted standard deviation of the values. The isotope shifts were determined relative to $^{221}$Fr, with the uncertainty propagated from the error of the fits, the scatter and the drift in centroid frequency of the $^{221}$Fr reference scans~\cite{Cocolios2013b}.

\subsection{Spectroscopic studies of $^{204}$Fr}

\begin{figure}
\includegraphics[width = \columnwidth]{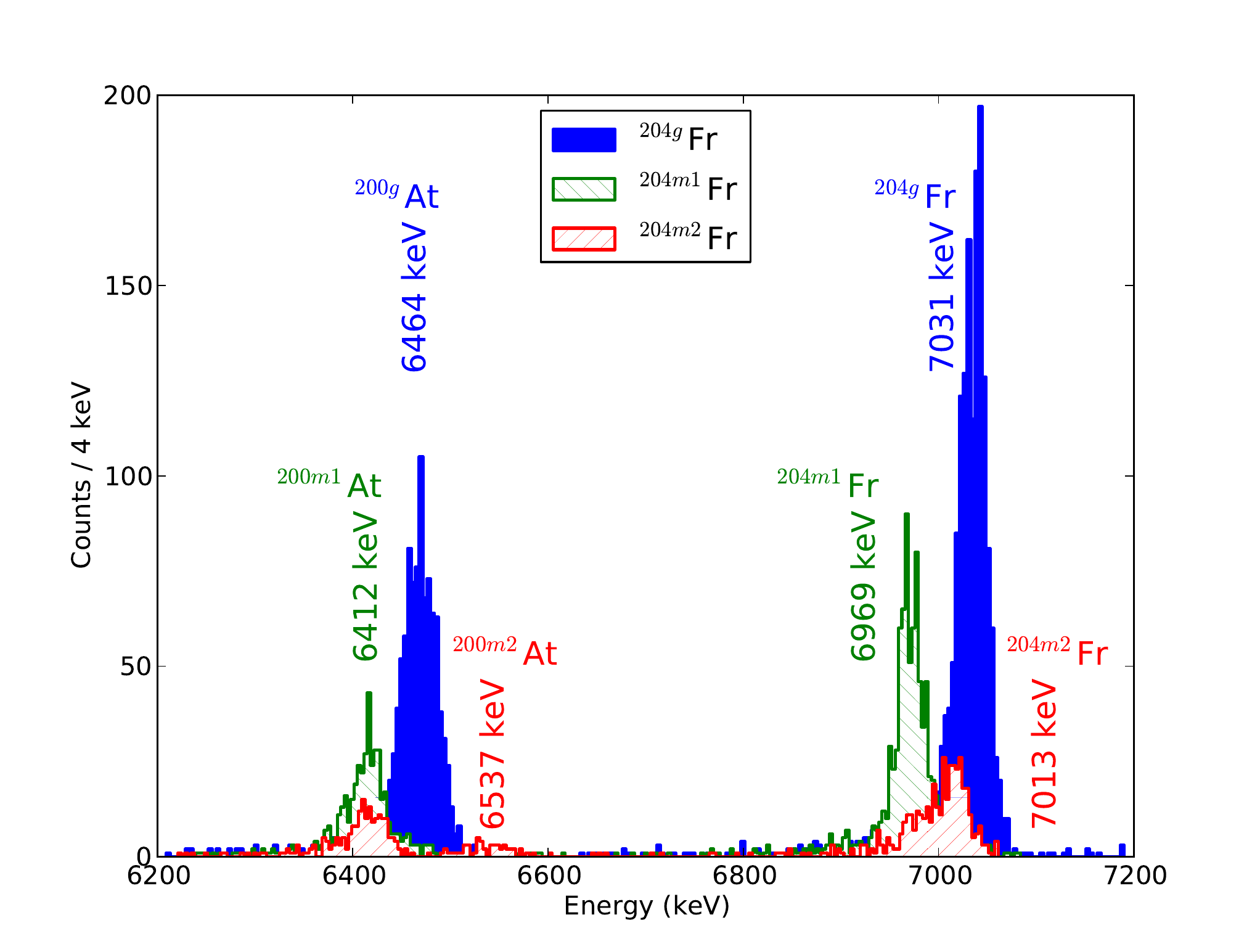}
\caption{\label{fig:DSS_Fr204_LANDS} Alpha-particle spectroscopy of (blue) $^{204g}$Fr, (green) $^{204m1}$Fr and (red) $^{204m2}$Fr allowed the hyperfine peaks in Fig.~\ref{fig:CRIS_Fr204} to be identified. The laser was detuned by 11.503~GHz (peak A, $^{204g}$Fr), 8.508~GHz (peak B, $^{204m1}$Fr) and 18.693~GHz (peak C, $^{204m2}$Fr) relative to the centroid frequency of $^{221}$Fr.}
\end{figure}

The hyperfine structure of $^{204}$Fr is shown in Fig.~\ref{fig:CRIS_Fr204}, measured by detecting the laser ions with the MCP detector as a function of the scanned first-step laser frequency. Five peaks are observed in the spectrum. Considering that only the lower-state splitting is resolved (associated with the 1.5~GHz linewidth of the scanning laser), two hyperfine resonances are expected per nuclear (ground or isomeric) state. Consequently, Fig.~\ref{fig:CRIS_Fr204} contains the hyperfine structure of three long-lived states in $^{204}$Fr, with one of the resonances unresolved (labeled E). In order to identify the states of the hyperfine resonances, laser assisted alpha-decay spectroscopy was used.

\begin{figure}
\includegraphics[width = \columnwidth]{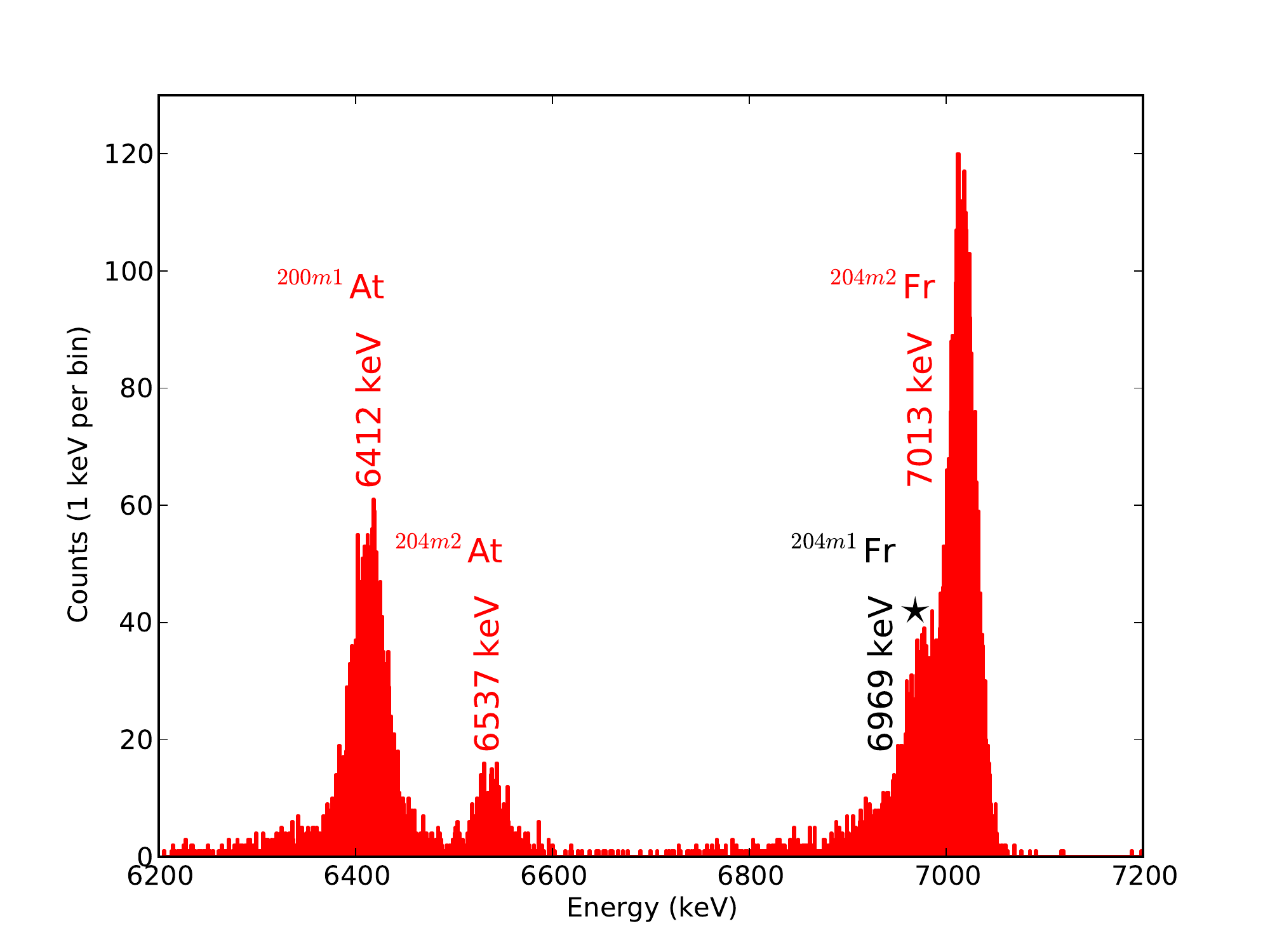}
\caption{\label{fig:DSS_Fr204m2_alpha} Alpha-particle spectroscopy of the (10$^-$) state of $^{204m2}$Fr. The decay of $^{204m2}$Fr to $^{204m1}$Fr via an E3 IT is observed through the presence of $^{204m1}$Fr alpha particles of 6969~keV (denoted by $\star$). The laser was detuned by 43.258~GHz relative to the centroid frequency of $^{221}$Fr.}
\end{figure}

The radioactive decay of the low-lying states in $^{204}$Fr is presented in Fig.~\ref{fig:Fr204_alpha_decay}. The characteristic alpha decay of each nuclear state in $^{204}$Fr was utilized to identify the hyperfine-structure resonances of Fig.~\ref{fig:CRIS_Fr204}. The laser was tuned on resonance with each of the first three hyperfine resonances (labeled A, B and C) and alpha-decay spectroscopy was performed on each. The alpha-particle energy spectrum of these three states is illustrated in Fig.~\ref{fig:DSS_Fr204_LANDS}. The energy of the alpha particles emitted when the laser was on resonance with an atomic transition of the hyperfine spectrum characteristic of $^{204g}$Fr is shown in blue. This transition occurred at 11.503~GHz (peak A of Fig.~\ref{fig:CRIS_Fr204}) relative to the centroid frequency of $^{221}$Fr. Similarly, the alpha-particle energy spectra for $^{204m1}$Fr and $^{204m2}$Fr are shown in green and red, when the laser was detuned by 8.508~GHz and 18.693~GHz (peak B and C) from the reference frequency, respectively. Present in the alpha-particle energy spectrum are the alpha particles emitted from the decay of the $^{204}$Fr states (6950-7050~keV) in addition to those emitted from the nuclear states in the daughter isotope $^{200}$At (6400-6500~keV). Each state in $^{204}$Fr has a characteristic alpha-particle emission energy: 7031~keV for $^{204g}$Fr, 6969~keV for $^{204m1}$Fr and 7013~keV for $^{204m2}$Fr. This was confirmed by the presence of the corresponding daughter decays of $^{200g}$At (6464~keV), $^{200m1}$Fr (6412~keV), and $^{200m2}$At (6537~keV) in the alpha-particle energy spectrum.

\begin{figure}
\includegraphics[width = \columnwidth]{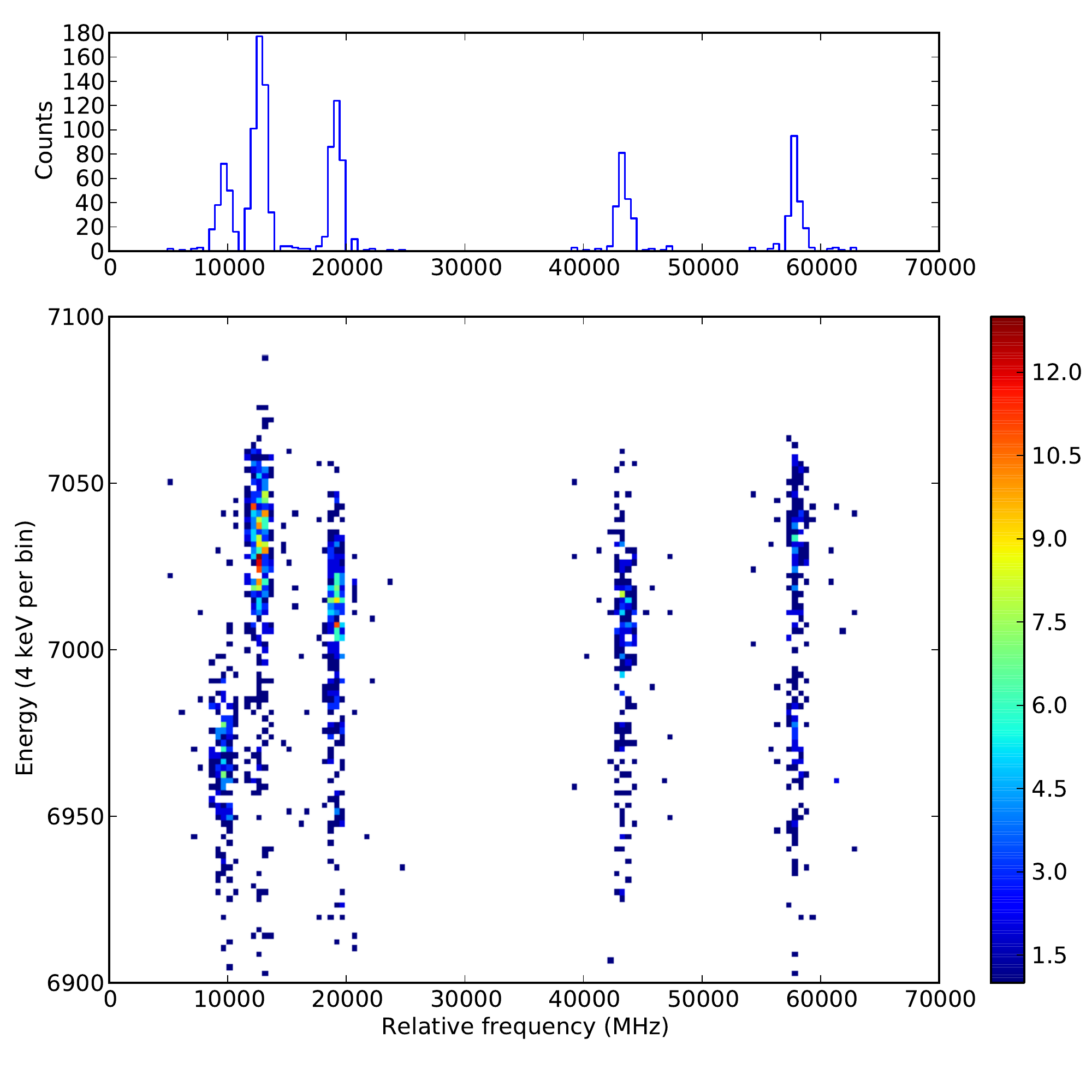}
\caption{\label{fig:DSS_Fr204_HFS_3D} A two-dimensional histogram of the alpha-particle energy as the hyperfine structure of $^{204}$Fr is probed. (Top) Projection of the frequency axis. The total number of alpha particles detected at each laser frequency reveals the hyperfine structure of $^{204}$Fr.}
\end{figure}

An additional alpha-decay measurement was performed on peak D in the hyperfine spectrum of $^{204}$Fr (see Fig.~\ref{fig:CRIS_Fr204}) at 43.258~GHz relative to the centroid frequency of $^{221}$Fr. The observation of 7013~keV alpha particles allowed this state to be identified as $^{204m2}$Fr. This meant the identity of all five hyperfine-structure peaks could be allocated to a state in $^{204}$Fr (hence the hyperfine structure peak E is the overlapping structure of $^{204g}$Fr and $^{204m1}$Fr), allowing analysis of the hyperfine structure of each state. 

In addition to the 7031~keV alpha particles of $^{204m2}$Fr, alpha particles of 6969~keV from the decay of $^{204m1}$Fr were also observed when the laser was on resonance with the $^{204m2}$Fr state. The decay of the (10$^-$) state to $^{204m1}$Fr via an E3 internal transition (IT) has been predicted~\cite{Huyse1992} but only recently observed~\cite{Jakobsson2012}, see Fig.~\ref{fig:Fr204_alpha_decay}. This was achieved by tagging the conversion electron from the internal conversion of $^{204m2}$Fr with the emitted 6969~keV alpha particles of $^{204m1}$Fr that followed (with a 5~s correlation time). This allowed the predicted energy of the 275~keV isomeric transition to be confirmed. During the CRIS experiment, an additional alpha-decay measurement was performed on $^{204m2}$Fr, with the laser detuned by 43.258~GHz relative to the centroid frequency of $^{221}$Fr, see Fig.~\ref{fig:DSS_Fr204m2_alpha}. This spectrum confirms the presence of the 6969~keV alpha particle (denoted by $\star$), emitted from the decay of $^{204m1}$Fr. The ultra-pure conditions of this measurement allowed the first unambiguous extraction of the branching ratios in the decay of $^{204m2}$Fr: $B_\alpha = 53(10)$\% and $B_{IT} = 47(10)$\%~\cite{MyThesis}.

\begin{figure}
\includegraphics[width = \columnwidth]{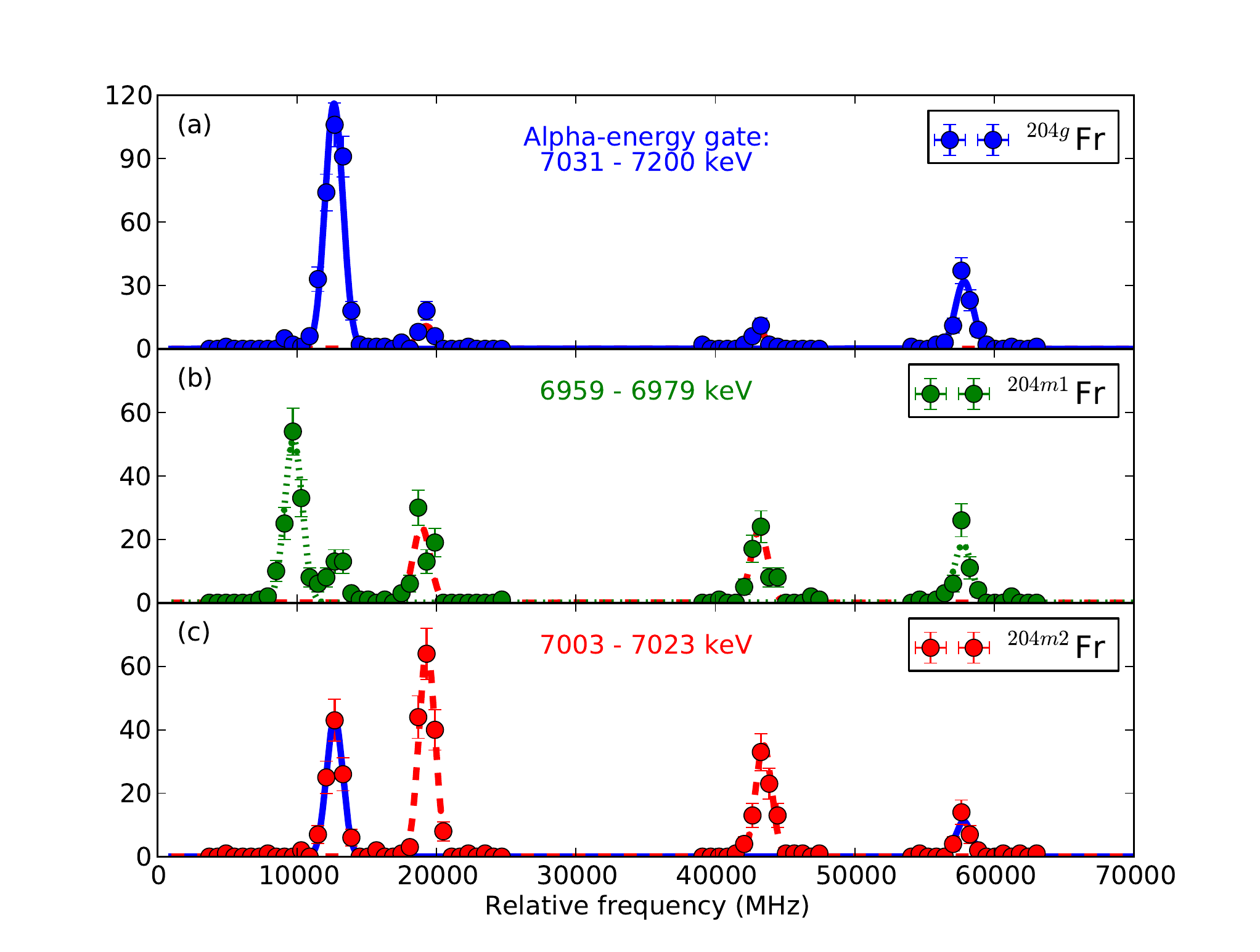}
\caption{\label{fig:DSS_Fr204_HFS} Alpha-tagged hyperfine structure of (a) $^{204g}$Fr spin 3$^{(+)}$ ground state, (b) $^{204m1}$Fr spin (7$^+$) isomer, and (c) $^{204m2}$Fr spin (10$^-$) isomer. Additional peaks are discussed in the text.}
\end{figure}

Decay-assisted laser spectroscopy was also performed on the hyperfine structure of the low-lying states of $^{204}$Fr. Just as the laser frequency of the resonant 422.7-nm ionization step was scanned and resonant ions were detected in the collinear resonance ionization spectroscopy of $^{204}$Fr, the same technique was repeated with the measurement of alpha particles. At each laser frequency, a radioactive-decay measurement of 60~s was made at the DSS, measuring the alpha particles emitted from the implanted ions. Fig.~\ref{fig:DSS_Fr204_HFS_3D}~(Top) shows the hyperfine peaks associated with each state in $^{204}$Fr. Measurement of the alpha decay as a function of laser frequency allowed production of a matrix of alpha-particle energy versus laser frequency, see Fig.~\ref{fig:DSS_Fr204_HFS_3D}.

In order to separate hyperfine structures for each states, an alpha-energy gating was used to maximize the signal-to-noise ratio for the alpha particle of interest. The alpha-energy gates were chosen to be 7031-7200~keV for $^{204g}$Fr, 6959-6979~keV for $^{204m1}$Fr and 7003-7023~keV for $^{204m2}$Fr.

By gating on the characteristic alpha-particle energies of the three states in $^{204}$Fr, the hyperfine structures of individual isomers become enhanced in the hyperfine spectrum. Fig.~\ref{fig:DSS_Fr204_HFS}(a) shows the hyperfine structure of $^{204g}$Fr, (b) $^{204m1}$Fr, and (c) $^{204m2}$Fr. The presence of $^{204m2}$Fr can be observed in the spectra of $^{204g}$Fr due to the overlapping peaks of the alpha energies: the tail of the 7013~keV alpha peak is present in the gate of the $^{204g}$Fr alpha peak. The presence of $^{204m2}$Fr in the hyperfine structure spectrum of $^{204m1}$Fr is attributed to the E3 IT decay of $^{204m2}$Fr to $^{204m1}$Fr: alpha particles of energy 6969~keV are observed when on resonance with $^{204m2}$Fr. Additionally, $^{204g}$Fr is present in the $^{204m2}$Fr spectra due to the similar energies of the 7031~keV and 7013~keV alpha particles. However, despite the contamination in the hyperfine spectra, each peak is separated sufficiently in frequency to be analysed independently.

From the resulting hyperfine structures of Fig.~\ref{fig:DSS_Fr204_HFS} produced by the alpha-tagging process (in comparison to the overlapping ion data of Fig.~\ref{fig:CRIS_Fr204}), each state of $^{204}$Fr can be analysed individually and the hyperfine factors extracted with better accuracy and reliability. The estimated error of the $A_{S_{1/2}}$ factors was 30~MHz on account of the scatter of $A_{S_{1/2}}$ values for $^{221}$Fr. Likewise, an error of 100~MHz was assigned to the isotope shifts.

\subsection{Identification of the hyperfine structure of $^{202}$Fr}

\begin{figure}
\includegraphics[width = \columnwidth]{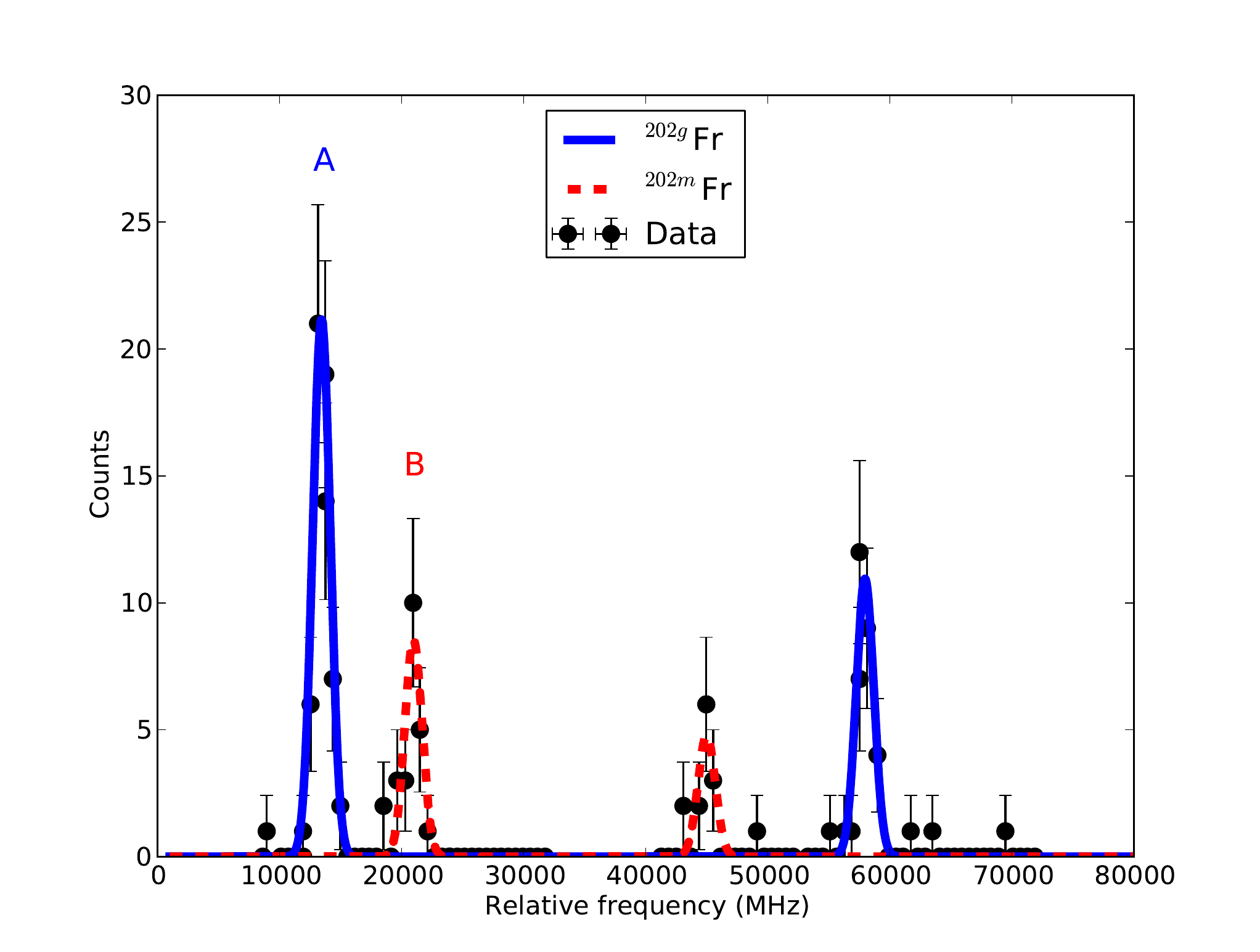}
\caption{\label{fig:CRIS_Fr202} Collinear resonance ionization spectroscopy of $^{202}$Fr relative to $^{221}$Fr. The hyperfine structure of the (3$^+$) ground state of $^{202g}$Fr is shown in blue and the (10$^-$) state of $^{202m}$Fr is shown in red.}
\end{figure}

\begin{figure}
\includegraphics[width = \columnwidth]{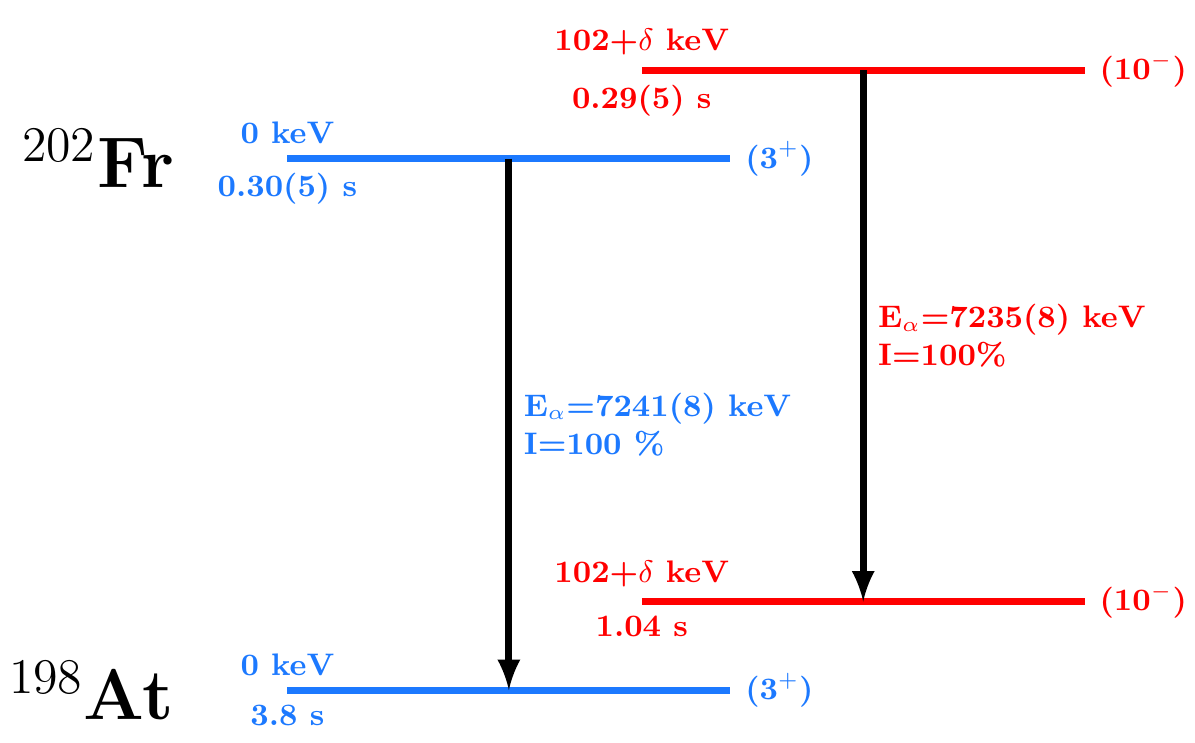}
\caption{\label{fig:Fr202_alpha_decay} The radioactive decay of $^{202}$Fr and its isomer~\cite{DeWitte2005}.}
\end{figure}

The hyperfine structure of $^{202}$Fr obtained with collinear resonance ionization spectroscopy is presented in Fig.~\ref{fig:CRIS_Fr202}. The four hyperfine resonances illustrate the presence of the ground ($^{202g}$Fr) and isomeric ($^{202m}$Fr) states. Identification of these two states was performed with laser-assisted alpha-decay spectroscopy. According to literature, the radioactive decay of $^{202g}$Fr (t$_{1/2}$ = 0.30(5)~s) emits an alpha particle of energy 7241(8)~keV, whereas $^{202m}$Fr (t$_{1/2}$ = 0.29(5)~s) emits an alpha particle of energy 7235(8)~keV~\cite{Zhu2008}. The radioactive decay of the ground and isomeric state of $^{202}$Fr is presented in Fig.~\ref{fig:Fr202_alpha_decay}. 

\begin{figure}
\includegraphics[width = \columnwidth]{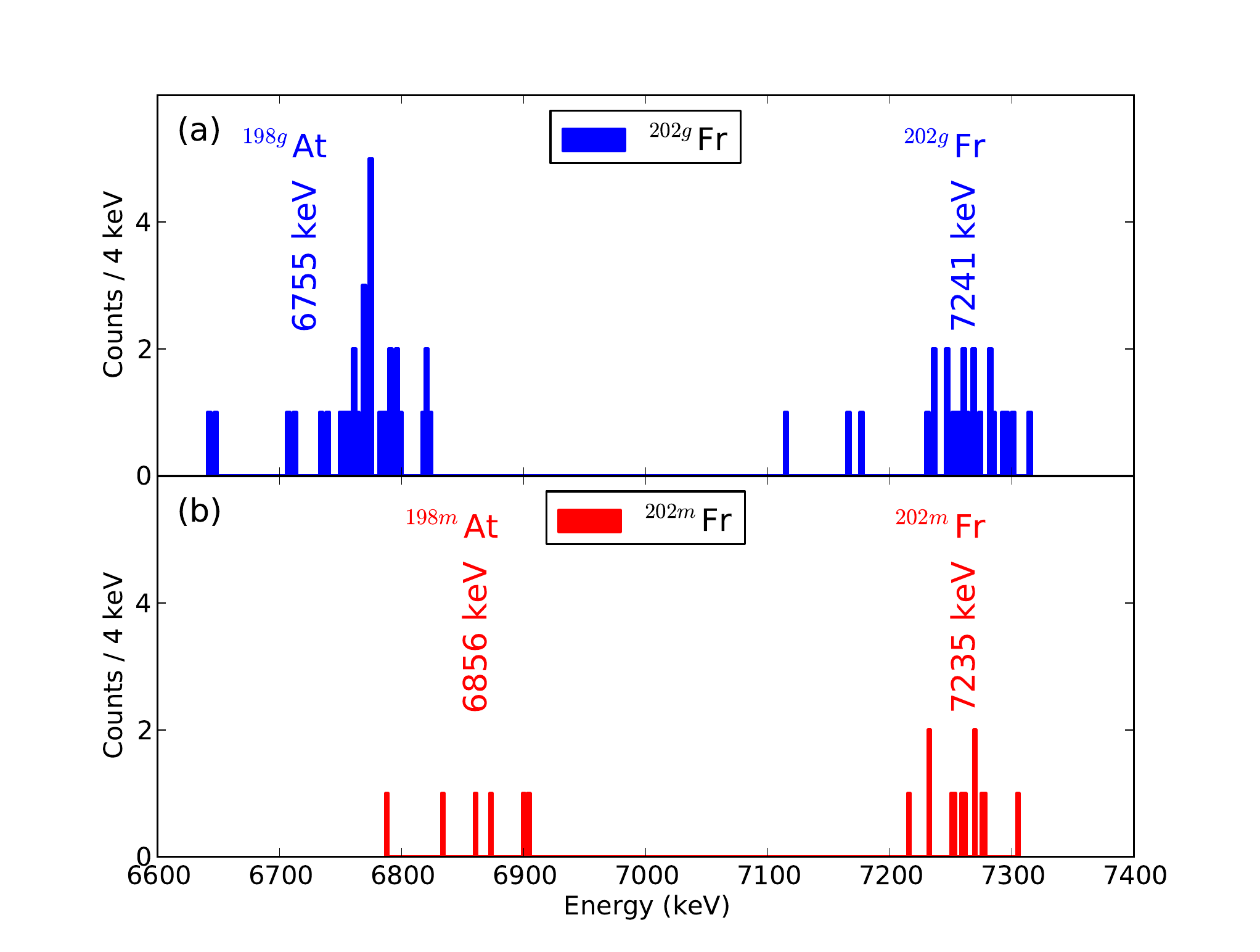}
\caption{\label{fig:DSS_Fr202_LANDS} Alpha-particle spectroscopy of (blue) $^{202g}$Fr and (red) $^{202m}$Fr allowed the hyperfine peaks in Fig.~\ref{fig:CRIS_Fr202} to be identified. The laser was detuned by (a) 13.760~GHz (peak A, $^{202g}$Fr) and (b) 20.950~GHz (peak B, $^{202m}$Fr) relative to the centroid frequency of $^{221}$Fr.}
\end{figure}

The laser was tuned onto resonance with peak A ($^{202g}$Fr, 13.760~GHz relative to the centroid frequency of $^{221}$Fr) and peak B ($^{202m}$Fr, 20.950~GHz relative to the centroid frequency of $^{221}$Fr) of Fig.~\ref{fig:CRIS_Fr202} obtained from ion detection. For each position, an alpha-decay measurement was performed, shown in Fig.~\ref{fig:DSS_Fr202_LANDS}. The alpha particles emitted when the laser was on resonance with an atomic transition characteristic to $^{202g}$Fr are shown in blue, and $^{202m}$Fr in red.  Due to the limited statistics of our measurement, and the similarity in energies of the alpha particles (within error), it is impossible to say that alpha particles of different energies are observed in Fig.~\ref{fig:DSS_Fr202_LANDS}.

Firm identification of the hyperfine components can be achieved however by studying the alpha particles emitted by the daughter isotopes $^{198g,m}$At. Evident in the spectrum of $^{202g}$Fr are the alpha particles emitted from the decay of the daughter nucleus $^{198g}$At with an energy of 6755~keV. Similarly, present in the $^{202m}$Fr spectrum are the alpha particles from the decay of $^{198m}$At with an energy of 6856~keV. The difference in energy of these two alpha peaks illustrates the ability of the CRIS technique to separate the two states and provide pure ground state and isomeric beams for decay spectroscopy.

\subsection{Isomer identification of the resonance spectrum of $^{206}$Fr}

\begin{figure}
\includegraphics[width = \columnwidth]{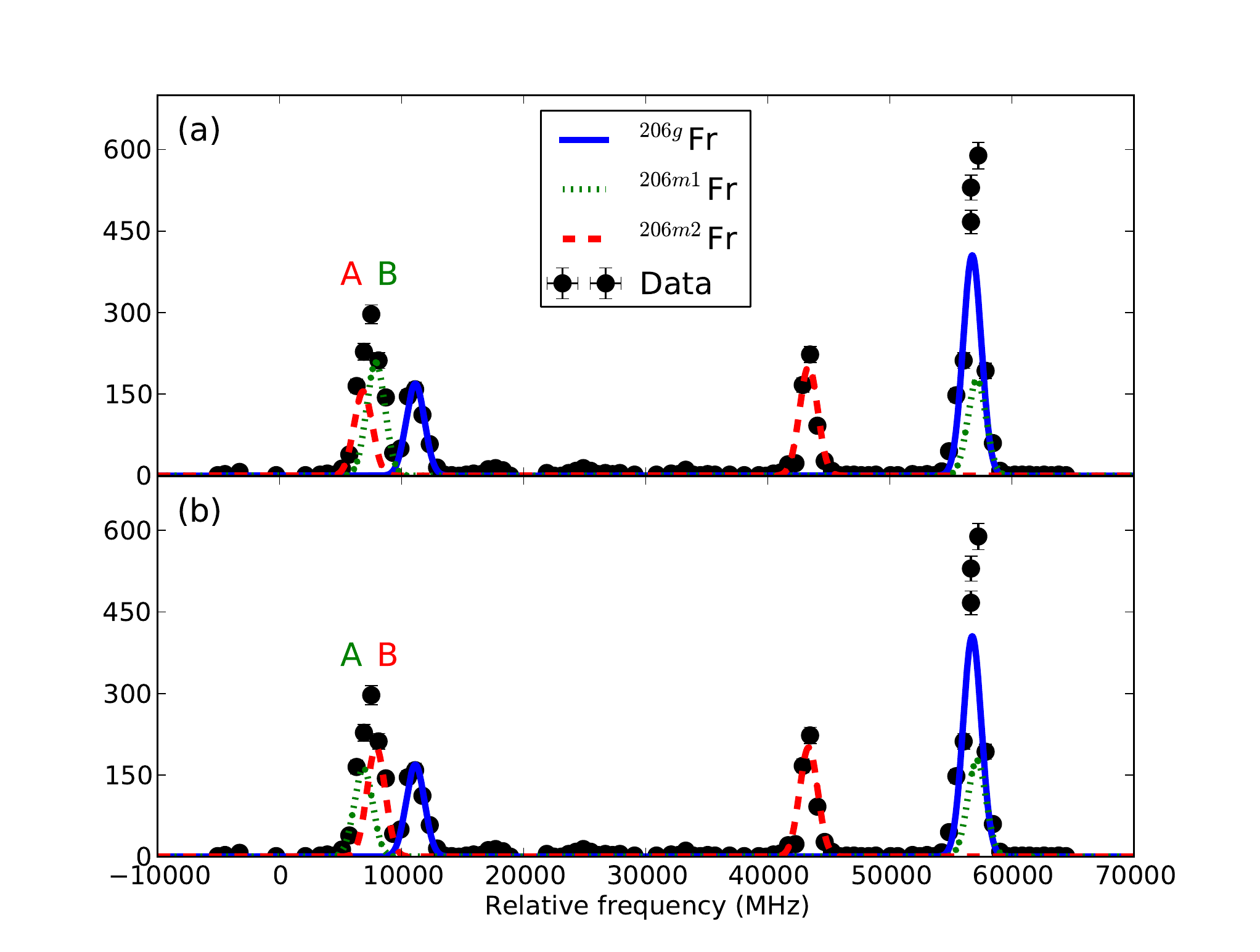}
\caption{\label{fig:CRIS_Fr206} Collinear resonance ionization spectroscopy of $^{206}$Fr relative to $^{221}$Fr. (a) Option 1: Peak A is assigned to $^{206m2}$Fr and peak B to $^{206m1}$Fr. (b) Option 2: Peak A is assigned to $^{206m1}$Fr and peak B to $^{206m2}$Fr. }
\end{figure}

Two sets of data were used in the determination of the nuclear observables from the hyperfine structure scans. The data for the francium isotopes $^{202-206,221}$Fr were taken in Run I and the data for $^{202-205,221}$Fr were taken in Run II. Consistency checks were carried out, allowing $^{206}$Fr to be evaluated with respect to the rest of the data set from Run II. A detailed description of this analysis can be found in Ref.~\cite{MyThesis}. In Run I, no alpha-tagging was available and consequently the peaks in the ion-detected hyperfine spectrum needed to be identified in a different way. Recent measurements of the ground-state hyperfine structure of $^{206g}$Fr provided the $A_{S_{1/2}}$ factor for the splitting of the 7s~$^2$S$_{1/2}$ state~\cite{Voss2013}. One peak of the (7$^+$) isomeric state was also identified in this experiment (see Fig.~1(c) of Ref.~\cite{Voss2013}), allowing the positions of the overlapping resonances to be determined. This left only the identity of peaks A and B (shown in Fig.~\ref{fig:CRIS_Fr206}) unknown. Fig.~\ref{fig:CRIS_Fr206}(a) presents the hyperfine structures when peak A is assigned to $^{206m2}$Fr and peak B to $^{206m1}$Fr. Fig.~\ref{fig:CRIS_Fr206}(b) shows the fit when peak A is $^{206m1}$Fr and peak B is $^{206m2}$Fr. The suggested identity of the two resonances (based on mean-square charge radii and $g$-factor systematics) is discussed in Sec.~\ref{sec:Discussion}.

\subsection{Yield measurements}

\begin{table}[h]
\caption{\label{tab:Yields}Yields of the neutron-deficient francium isotopes at the ISOLDE facility (1.4~GeV protons on a UC$_x$ target). The nuclear-state composition of the radioactive beams for $^{202,204,206}$Fr are presented.}
\begin{ruledtabular}
\begin{tabular}{l c r r r } 
$A$ 		& Yield				& \multicolumn{3}{c}{Proportion of beam}			\\ 
		&  (ions/s)				& Spin 3$^{(+)}$ 	& Spin (7$^+$) 		& Spin (10$^-$) \\ \hline
202		& $1 \times 10^2$		& 76(14)\%		& 				& 24(6)\% 	\\  
203 		& $1 \times 10^3$ 		& 				& 				& 			\\ 
204		& $1 \times 10^4$\footnote{Estimate based on yield systematics.} 	& 63(3)\%			& 27(3)\%			& 10(1)\% 	\\  
205		& $2 \times 10^5$ 		&  				& 			 	& 			\\ 
206		& $3 \times 10^6$ 		& 63(7)\%			& 27(5)\%			& 9(1)\% 	\\  
\end{tabular}
\end{ruledtabular}
\end{table}

The yields of the neutron-deficient francium isotopes $^{202-206}$Fr are presented in Table~\ref{tab:Yields}. The quoted yields, scaled ISOLDE-database yields based on an independent yield measurement of $^{202}$Fr, can be expected to vary by a factor of two due to different targets. The quoted value for $^{204}$Fr is estimated based on francium yields systematics. The composition of the beam for $^{202,204,206}$Fr was calculated from the ratio of hyperfine-peak intensities (based on the strongest hyperfine-structure resonance) from the CRIS ion data. The composition of the beam for $^{202}$Fr was confirmed with the alpha-decay data measured with the DSS.

\subsection{King-plot analysis}

\begin{figure}
\includegraphics[width = \columnwidth]{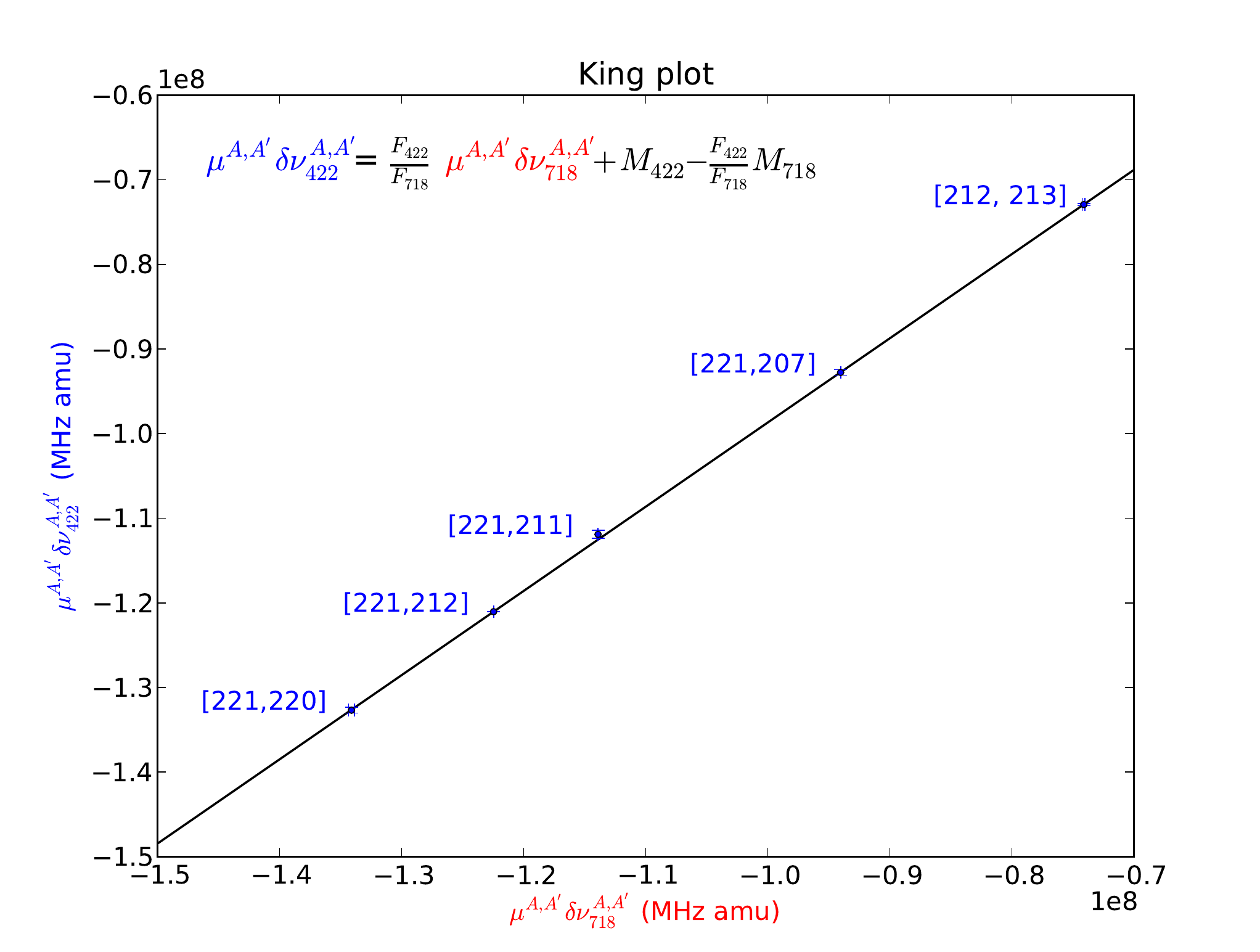}
\caption{\label{fig:CRIS_Fr_King_plot} A King plot for the extraction of atomic factors $F$ and $M$ for the 422.7-nm transition. See text for details.}
\end{figure}

\begin{table*}
\caption{\label{tab:CRIS_HFS_Fr}Spins, half-lives, $A_{S_{1/2}}$ factors, isotope shifts, magnetic moments and change in mean-square charge radii of the neutron-deficient francium isotopes $^{202-206}$Fr with reference to $^{221}$Fr for the 7s~$^2$S$_{1/2}$ $\rightarrow$ 8p~$^2$P$_{3/2}$ atomic transition. All $A_{S_{1/2}}$-factor and magnetic-moment values were deduced using the nuclear spins presented. The half-life values are taken from Refs.~\cite{LourensThesis, Ritchie1981, Huyse1992, Uusitalo2005, Kondev2011, Singh2013, Browne2011}.}
\begin{ruledtabular}
\begin{tabular}{l r r r r r r r r r r } 
$A$ 			&$I$ 			& $t_{1/2}$ (s) & \multicolumn{2}{c}{$A_{S_{1/2}}$ (GHz)} & \multicolumn{2}{c}{$\mu$ ($\mu_N$)} & \multicolumn{1}{c}{$\delta \nu^{A,221}$ (GHz)} & \multicolumn{2}{c}{$\delta \langle r^2 \rangle$ (fm$^2$)}  \\
			& 				& \multicolumn{1}{c}{Lit.} 		&\multicolumn{1}{c}{Exp.} & \multicolumn{1}{c}{Lit.} 	& \multicolumn{1}{c}{Exp.}  & \multicolumn{1}{c}{Lit.} & \multicolumn{1}{c}{Exp.}  	& \multicolumn{1}{c}{Exp.} & \multicolumn{1}{c}{Lit.}	\\ \hline
202g 		& (3$^+$) 			& 0.30(5)		& +12.80(5) 	&	 	& +3.90(5)	& & 32.68(10)  	& -1.596(18) 	 \\  
202m 		& (10$^-$) 		& 0.29(5)		& +2.30(3) 	&		& +2.34(4)	& & 32.57(13) 	& -1.591(19) 	 \\  
203 			& (9/2$^-$) 		& 0.53(2)		& +8.18(3) 	&		& +3.73(4)  	& & 31.32(10)	& -1.530(18)	\\ 
204g\footnote{Calculated from the alpha-decay gated hyperfine structure scan of $^{204}$Fr. See text for details.} 	& 3$^{(+)}$ 	& 1.9(5)	& +12.99(3) & +13.1499(43)\footnote{Literature value taken from Ref.~\cite{Voss2013}.} & +3.95(5) & +4.00(5)\footnotemark[2]\textsuperscript{,}\footnote{Literature magnetic-moment values re-calculated in reference to $\mu$($^{210}$Fr)~\cite{Gomez2008}}	& 32.19(10)  & -1.571(18) & -1.5542(4)\footnotemark[2]	 \\
204m1\footnotemark[1]	& (7$^+$) 	& 1.6(5)		&+6.44(3) 	&		 & +4.57(6)	& & 32.32(10)	& -1.577(18) 	 \\ 
204m2\footnotemark	[1]	& (10$^-$)& 0.8(2)		& +2.31(3) 	&		 & +2.35(4)	& & 30.99(10)	& -1.513(17) 	 \\  
205			& 9/2$^-$ 			& 3.96(4)		& +8.40(3) 	& +8.3550(11)\footnotemark[2]		& +3.83(5) 	&+3.81(4)\footnotemark[2]\textsuperscript{,}\footnotemark[3]	& 30.21(10) 	& -1.475(17) & -1.4745(4)\footnotemark[2] 	\\ 
206g			& 3$^{(+)}$ 		& 15.9(3)		& +13.12(3) 	& +13.0522(20)\footnotemark[2]		& +3.99(5)	&+3.97(5)\footnotemark[2]\textsuperscript{,}\footnotemark[3]	& 30.04(12) 		& -1.465(17) & -1.4768(4)\footnotemark[2] 	 \\ 
206m1\footnote{Based on the isomeric identity of the hyperfine resonances of Option 1. See text for details.} 		& (7$^+$) 			& 15.9(3)	& +6.61(3) &	 & +4.69(6) &	& 30.23(16)	& -1.475(18) &	\\  
206m2\footnotemark[4] & (10$^-$) 	& 0.7(1)		& +3.50(3) 	&	 	& +3.55(5)	&  & 23.57(12)	& -1.153(14) &	 \\
206m1\footnote{Based on the isomeric identity of the hyperfine resonances of Option 2. See text for details.}		& (7$^+$) 			& 15.9(3)	& +6.74(4)  & 	 & +4.79(6) &	& 29.69(15)	& -1.449(17) &	\\  
206m2\footnotemark[5] & (10$^-$) 	& 0.7(1)		& +3.40(3) 	&		 & +3.45(5) 	& & 24.13(12)	& -1.180(14) 	& \\
207		 	& 9/2$^-$			& 14.8(1) 		& +8.48(3) 	& +8.484(1)\footnote{Literature value taken from Ref.~\cite{Coc1985}.} 		 & +3.87(5)	&+3.87(4)\footnotemark[3]\textsuperscript{,}\footnotemark[6]		& 28.42(10)	& -1.386(16) 	& -1.386(3)\footnote{Literature value taken from Ref.~\cite{Dzuba2005}.} 	 \\  
211		 	& 9/2$^-$	 		& 186(1)		& +8.70(6) 	& +8.7139(8)\footnotemark[6]	& +3.97(5)	&+3.98(5)\footnotemark[3]\textsuperscript{,}\footnotemark[6]  & 24.04(10) 	& -1.171(13) 	& -1.1779(4)\footnotemark[7] 	 \\  
220		 	& 1$^+$			& 27.4(3)		& -6.50(4) 	& -6.5494(9)\footnotemark[6]	& -0.66(1)		&-0.66(1)\footnotemark[3]\textsuperscript{,}\footnotemark[6] & 2.75(10) 		& -0.134(5) 	& -0.133(10)\footnotemark[7]	 \\  
221		 	& 5/2$^-$ 			& 294(12)		& +6.20(3) 	& +6.2046(8)\footnotemark[6]	& +1.57(2)	&+1.57(2)\footnotemark[3]\textsuperscript{,}\footnotemark[6] & 0 	& 0			& 	 \\  
\end{tabular}
\end{ruledtabular}
\end{table*}

The atomic factors $F$ and $M$ were evaluated by the King-plot method~\cite{King1963}. This combines the previously measured isotope shifts by Coc~\cite{Coc1985} of the 7s~$^2$S$_{1/2}$ $\rightarrow$ 7p~$^2$P$_{3/2}$ transition with 718-nm laser light, with those made by Duong~\cite{Duong1987} of the 7s~$^2$S$_{1/2}$ $\rightarrow$ 8p~$^2$P$_{3/2}$ transition (422.7~nm). The isotope shifts of $\delta \nu^{207,221}$ and $\delta \nu^{211,221}$ from this work were combined with $\delta \nu^{220,221}$ and $\delta \nu^{213,212}$ from Duong. These values were plotted against the corresponding isotope shifts from Coc~\cite{Coc1985}, shown in Fig.~\ref{fig:CRIS_Fr_King_plot}. From the linear fit of the data, and using
\begin{equation} \label{eq:KP} \mu^{A,A'} \delta \nu_{422}^{A,A'} = \frac{F_{422}}{F_{718}}\mu^{A,A'}\delta \nu_{718}^{A,A'} + M_{422} - \frac{F_{422}}{F_{718}}M_{718} ,\end{equation}
where $\mu^{A,A'}=AA'/(A'-A)$, enabled the evaluation of $F_{422}/F_{718} = +0.995(3)$ and $M_{422} - (F_{422}/F_{718})M_{718} =  +837(308)~\text{GHz~amu}$ respectively. From these values, the atomic factors for the 422.7-nm transition were calculated to be
\[ F_{422} = -20.67(21) \text{~GHz/fm}^2, \]
\[ M_{422} = +750(330) \text{~GHz~amu}. \]
For comparison, the atomic factors evaluated for the 718-nm transition were determined by Dzuba to be $F_{718} = -$20.766(208)~GHz/fm$^2$ and $M_{718} = -$85(113)~GHz~amu~\cite{Dzuba2005}. 

The mass factor is the linear combination of two components: the normal mass shift, $K^{\textnormal{NMS}}$, and the specific mass shift, $K^{\textnormal{SMS}}$,
\begin{equation} M_{422} = K^{\textnormal{NMS}}_{422} + K^{\textnormal{SMS}}_{422}, \end{equation}
and is dependent on the frequency of the transition probed. Subtraction of the normal mass shift of the 422.7-nm transition ($K^{\textnormal{NMS}}_{422} = +389$~GHz amu) from the mass factor $M_{422}$ allows for calculation of the specific mass shift, giving $K^{\textnormal{SMS}}_{422} = +360(330)$~GHz~amu. The specific mass shift for the 718-nm line was determined by Dzuba to be $K^{\textnormal{SMS}}_{718} = -314(113)$~GHz~amu~\cite{Dzuba2005}.

\subsection{Hyperfine structure observables}

Table~\ref{tab:CRIS_HFS_Fr} presents the hyperfine $A_{S_{1/2}}$ factor, isotope shift, change in mean-square charge radius and magnetic moment values extracted from the CRIS data for the francium isotopes $^{202-206}$Fr with reference to $^{221}$Fr. Additional data for $^{207,211,220}$Fr (used in the creation of the King plot of Fig.~\ref{fig:CRIS_Fr_King_plot}) in included for completeness. All values were deduced using the nuclear spins presented. 

The hyperfine $A_{S_{1/2}}$ factor is defined as
\begin{equation} \label{eq:A} A = \frac{\mu_IB_e}{I\cdot J} ,\end{equation}
with $\mu_I$ the magnetic dipole moment of the nucleus and $B_e$ the magnetic field of the electrons at the nucleus. For each isotope, it was calculated from the weighted mean of $A_{S_{1/2}}$ values for isotopes where more than one hyperfine structure scan is present. A minimum error of 30~MHz was attributed to the $A_{S_{1/2}}$ factor values due to the scatter of the measured $A_{S_{1/2}}$ for the reference isotope $^{221}$Fr~\cite{Cocolios2013b, Lynch2013b}. 

The isotope shift, $\delta \nu^{A,A'}$, between isotopes $A$ and $A'$ is expressed as
\begin{equation}\label{eq:IS} \delta \nu^{A,A'} = M \frac{A'-A}{AA'} + F \delta \langle r^2 \rangle^{A,A'} .\end{equation} 
As with the $A_{S_{1/2}}$ values, the isotope shifts were calculated as the weighted mean of all isotope shifts for a given nucleus. The error on the isotope shift was determined to be 100~MHz due to the long-term drift of the centroid frequency of $^{221}$Fr as the experiment progressed, and the scan-to-scan scatter in centroid frequency. When the calculated weighted standard deviation of the isotope shift was higher than 100~MHz, this error is quoted instead. Combining the extracted $F$ and $M$ atomic factors from the King-plot analysis with the measured isotope shifts, evaluation of the change in mean-square charge radii, $\delta \langle r^2 \rangle^{A,A'}$, between francium isotopes can be performed, see Eq.~(\ref{eq:IS}).

The magnetic moment of the isotopes under investigation can be extracted from the known moment of another isotope of the element, using the ratio
\begin{equation} \label{eq:mu} \mu = \mu_{ref} \frac{IA}{I_{ref}A_{ref}} .\end{equation}
In this work, calculation of the magnetic moments was evaluated in reference to the magnetic moment of $^{210}$Fr, measured by Gomez ($\mu=+4.38(5)~\mu_N$, $I^\pi = 6^+$, $A_{S_{1/2}}=+7195.1(4)$~MHz~\cite{Gomez2008, Coc1985}). This represents the most accurate measurement of the magnetic moment of a francium isotope to date, due to probing the 9s~$^2$S$_{1/2}$ hyperfine splitting which has reduced electron-correlation effects than that of the ground state. The current evaluated magnetic moments of the francium isotopes are made in reference to the magnetic moment of $^{211}$Fr of Ekstr\"om~\cite{Ekstrom1986}.

The hyperfine anomaly for the francium isotopes is generally considered to be of the order of 1\% and is included as a contribution to the error of the hyperfine $A_{S_{1/2}}$ factors and magnetic moments~\cite{Stroke1961}.

Table~\ref{tab:CRIS_HFS_Fr} presents the experimental results alongside comparison to literature of the hyperfine $A_{S_{1/2}}$ factor, change in mean-square charge radius and magnetic moment values. The literature values for $^{204-206}$Fr have been taken from Ref.~\cite{Voss2013} and $^{207,211,220,221}$Fr from Ref.~\cite{Coc1985}. The magnetic-moment values from literature have been re-calculated in reference to $\mu$($^{210}$Fr)~\cite{Gomez2008}, the most accurate measurement to date. The change in mean-square charge radii values for $^{207,211,220}$Fr have been taken from Ref.~\cite{Dzuba2005}. All experimental results are in broad agreement with those of literature.

\section{\label{sec:Discussion} Discussion}

\subsection{Charge radii of the neutron-deficient francium}

\begin{figure}
\includegraphics[width = \columnwidth]{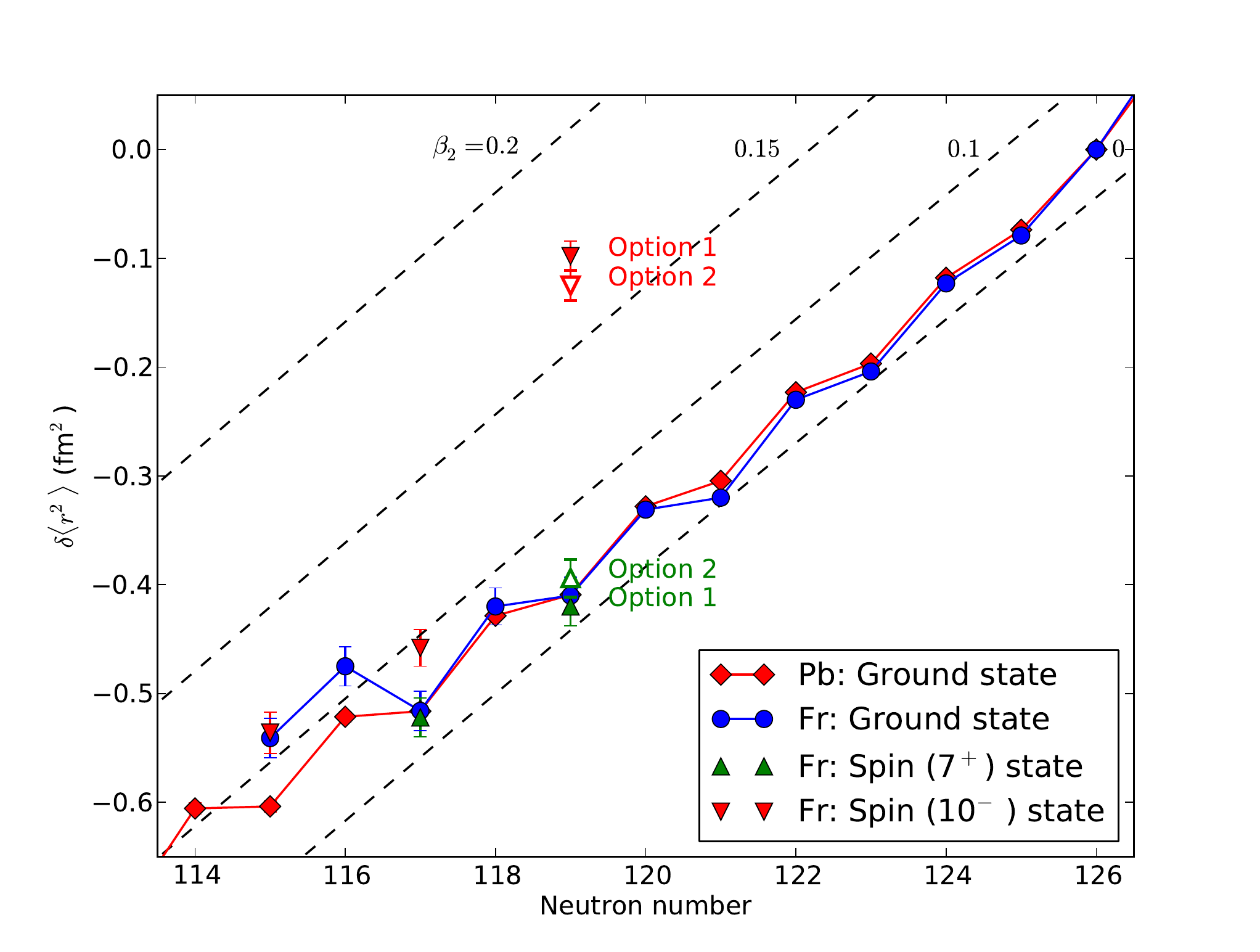}
\caption{\label{fig:CRIS_CR_Pb_Fr} Mean-square charge radii of the francium (circle) isotopes~\cite{Ekstrom1986} presented alongside the lead (diamond) isotopes~\cite{Anselment1986}. The dashed lines represent the prediction of the droplet model for given iso-deformation~\cite{Myers1983}. The data were calibrated by using $\beta_2$($^{213}$Fr)$=0.062$, evaluated from the energy of the $2^+_1$ state in $^{212}$Rn~\cite{Raman2001}. Option 1 and 2 for the (7$^+$) and (10$^-$) states in $^{206}$Fr are based on the isomeric identification given in Fig.~\ref{fig:CRIS_Fr206}.}
\end{figure}

Located between radon and radium, francium (Z~=~87) has 5 valence protons occupying the $\pi$1h$_{9/2}$ orbital, according to the shell model of spherical nuclei. Below the N~=~126 shell closure, the neutron-deficient francium isotopes were studied down to $^{202}$Fr (N~=~115). The change in mean-square charge radii for the francium and lead isotopes are presented in Fig.~\ref{fig:CRIS_CR_Pb_Fr}. The data of francium show the charge radii of $^{207-213}$Fr re-evaluated by Dzuba~\cite{Dzuba2005} alongside the CRIS values which extends the data set to $^{202}$Fr. The blue data points show the francium ground states, while the (7$^+$) isomeric states are in green and the (10$^-$) states in red. The error bars attributed to the CRIS values are propagated from the experimental error of the isotope shift and the systematic error associated with the atomic factors $F_{422}$ and $M_{422}$. The systematic error is the most significant contribution to the uncertainty associated with the mean-square charge radii, and not that arising from the isotope shift. The francium data is presented with the lead data of Anselment~\cite{Anselment1986} to illustrate the departure from the spherical nucleus. The change in mean-square charge radii of the francium isotopes have been overlapped with the charge radii of the lead isotopes, by using $^{213}$Fr (N~=~126) and $^{208}$Pb (N~=~126) as reference points. The dashed iso-deformation lines represent the prediction of the droplet model for the francium isotopes~\cite{Myers1983}. The data were calibrated using $\beta_2$($^{213}$Fr)$=0.062$, evaluated from the energy of the $2^+_1$ state in $^{212}$Rn~\cite{Raman2001}.

The doubly-magic $^{208}$Pb represents a model spherical nucleus, with the shape of the nucleus remaining spherical with the removal of neutrons from the closed N~=~126 shell. This trend is observed until N~=~114, where a small deviation from the spherical droplet model (isodeformation line $\beta_2=0.0$) is interpreted as enhanced collectivity due to the influence of particle-hole excitations across the Z~=~82 shell closure~\cite{DeWitte2007}. The change in mean-square charge radii for the francium isotopes shows agreement with the lead data as the $\nu$3p$_{3/2}$, $\nu$2f$_{5/2}$ and $\nu$3p$_{1/2}$ orbitals are progressively depleted. The deviation from sphericity at N~=~116 with $^{203}$Fr marks the onset of collective behaviour. The spectroscopic quadrupole moments were not measured in this work, since they require a laser linewidth of $<$100~MHz. Measurement of the quadrupole moment will provide information on the static deformation component of the change in mean-square charge radii, allowing better understanding of this transition region.

\begin{figure}
\includegraphics[width = \columnwidth]{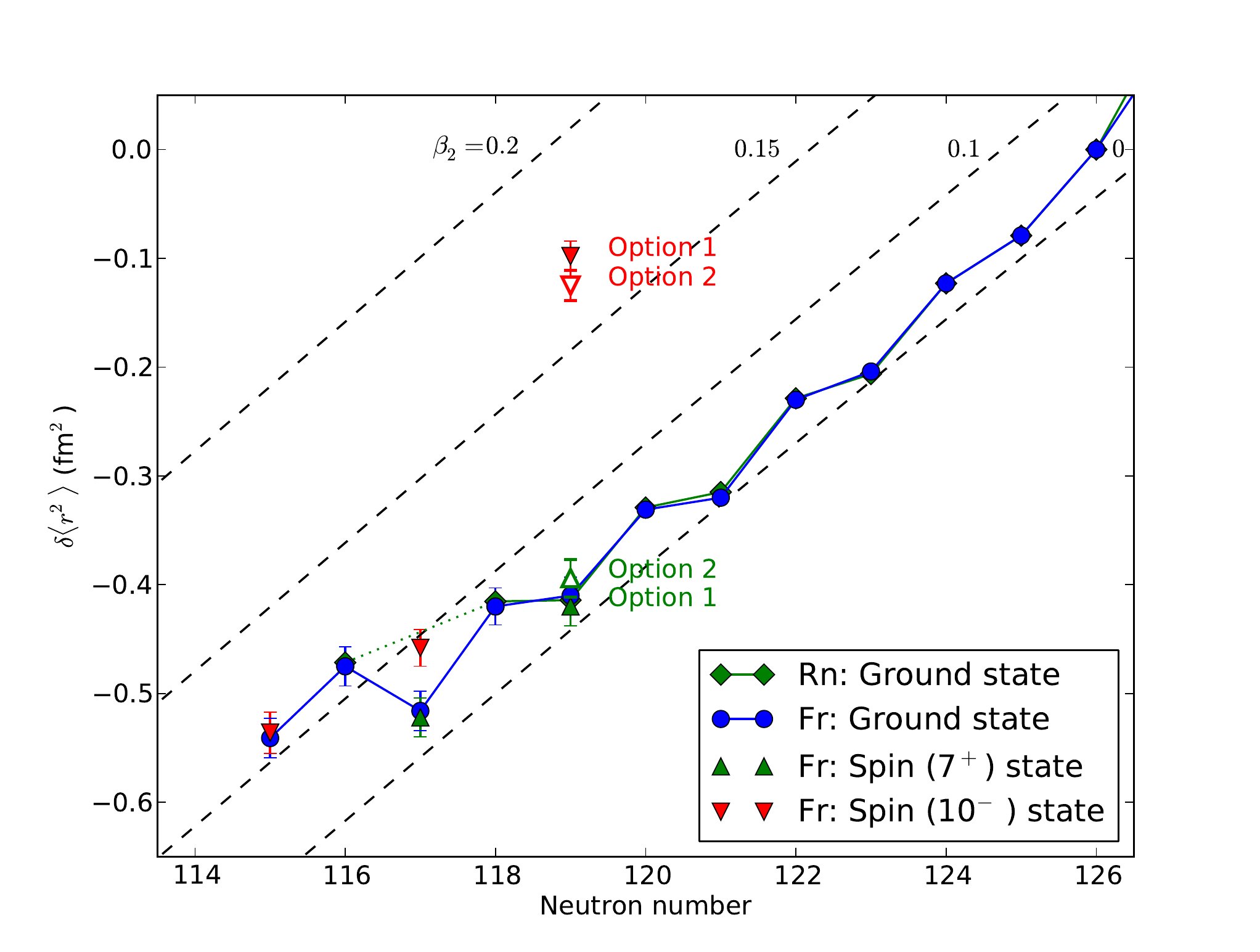}
\caption{\label{fig:CRIS_CR_Rn_Fr} Mean-square charge radii of the francium (circle) isotopes~\cite{Ekstrom1986} presented alongside the radon (diamond) isotopes~\cite{Borchers1987}. The dashed lines represent the prediction of the droplet model for given iso-deformation~\cite{Myers1983}. The data were calibrated using $\beta_2$($^{213}$Fr)$=0.062$, evaluated from the energy of the $2^+_1$ state in $^{212}$Rn~\cite{Raman2001}. Option 1 and 2 for the (7$^+$) and (10$^-$) states in $^{206}$Fr are based on the isomeric identification given in Fig.~\ref{fig:CRIS_Fr206}.}
\end{figure}

Recent laser spectroscopy measurements on the ground-state properties of $^{204,205,206}$Fr suggest this deviation occurs earlier, at $^{206}$Fr (N~=~119)~\cite{Voss2013}. In Ref.~\cite{Voss2013}, a more pronounced odd-even staggering is observed in relation to the lead isotopes, where the mean-square charge radius of $^{205}$Fr is larger than that of $^{206}$Fr. The CRIS experiment observed a smaller mean-square charge radius of $^{205}$Fr in comparison to $^{206}$Fr, the deviation from the lead isotopes occurring at $^{203}$Fr instead. However, both experiments are in broad agreement within errors down to N~=~117.

Figure~\ref{fig:CRIS_CR_Pb_Fr} presents the two options of the mean-square charge radii of $^{206m1}$Fr and $^{206m2}$Fr (as defined by their hyperfine peak identity in Fig.~\ref{fig:CRIS_Fr206}). Option 1 is favoured over option 2 due to the smaller mean-square charge radii of $^{206m1}$Fr (compared to $^{206g}$Fr) agreeing with the systematics of the states in $^{204}$Fr. As seen in Fig.~\ref{fig:CRIS_CR_Pb_Fr}, $^{206g}$Fr (N~=~119) overlaps with the lead data within errors. The large change in the mean-square charge radius of $^{206m2}$Fr suggests a highly deformed state for the (10$^-$) isomer.

The mean-square charge radii of francium are overlaid with the radon (Z~=~86) charge-radii of Borchers (down to N~=~116, with the exception of N~=~117)~\cite{Borchers1987} in Fig.~\ref{fig:CRIS_CR_Rn_Fr}. The mean-square charge-radii of radon have been calibrated to the francium pair $\delta \langle r^2 \rangle^{211,213}$ to account for the uncertainty in $F$ and $M$ for the optical transition probed (the original isotope shifts are presented graphically). Despite this, the agreement between the mean-square charge radii of the francium and radon data is clear. The addition of a single $\pi$1h$_{9/2}$ proton outside the radon even-Z core does not affect the charge-radii trend, suggesting the valence proton acts as a spectator particle.

Table~\ref{tab:Beta2} presents a comparison of $\beta_2$ values with literature. The droplet model~\cite{Myers1983} was used to extract the rms values for $\beta_2$ (column 3) from the change in mean-square charge radii (calibrated using $\beta_2$($^{213}$Fr)$=0.062$, as before). Column 4 presents $\beta_2$ values extracted from the quadrupole moments of Ref.~\cite{Voss2013}. The larger $\beta_2$ values extracted from the mean-square charge radii, compared to those extracted from the quadrupole moments, suggest that the enhanced collectivity observed in Figs.~\ref{fig:CRIS_CR_Pb_Fr} and~\ref{fig:CRIS_CR_Rn_Fr} is due to a large dynamic component of the nuclear deformation.

\begin{table}[t]
\caption{\label{tab:Beta2}Extracted $\beta_2$ values. (Exp.) The droplet model~\cite{Myers1983} was used to extract the rms values for $\beta_2$ from the change in mean-square charge radii. The charge-radii values were calibrated using $\beta_2$($^{213}$Fr)$=0.062$, as before. (Lit.) $\beta_2$ values were extracted from the quadrupole moments of Ref.~\cite{Voss2013}. See text for details.}
\begin{ruledtabular}
\begin{tabular}{l c c c} 
$A$ 		& $I$			& $\langle \beta_2^2 \rangle^{1/2}$ & $\beta_2$	\\ 
		&  			& Exp.		& Lit. 	\\ \hline
202g		& (3$^+$)		&  0.11 		& 		\\  
202m	& (10$^-$)	&  0.11 		& 		\\  
203 		& (9/2$^-$)	&  0.11		& 		\\ 
204g		& 3$^{(+)}$	&  0.06\footnote{Calculated from the alpha-decay gated hyperfine structure scan of $^{204}$Fr. See text for details.}	& -0.0140(14)	\\  
204m1	& (7$^+$)		&  0.06\footnotemark[1]		&		\\  
204m2	& (10$^-$)	&  0.09\footnotemark[1]		& 		\\  
205		& 9/2$^-$		&  0.08 		& -0.0204(2) 	\\ 
206g		& 3$^{(+)}$	&  0.05 		& -0.0269(8)	\\  
206m1	& (7$^+$)		&  0.04\footnote{Based on the isomeric identity of the hyperfine resonances of Option 1. See text for details.}			& 		\\  
206m2	& (10$^-$)	&  0.17\footnotemark[2]		& 		\\  
202m1	& (7$^+$)		&  0.07\footnote{Based on the isomeric identity of the hyperfine resonances of Option 2. See text for details.}			& 		\\  
202m2	& (10$^-$)	&  0.17\footnotemark[3]		& 		\\  
\end{tabular}
\end{ruledtabular}
\end{table}

\subsection{Interpretation of the nuclear $g$-factors}

\begin{figure}
\includegraphics[width = \columnwidth]{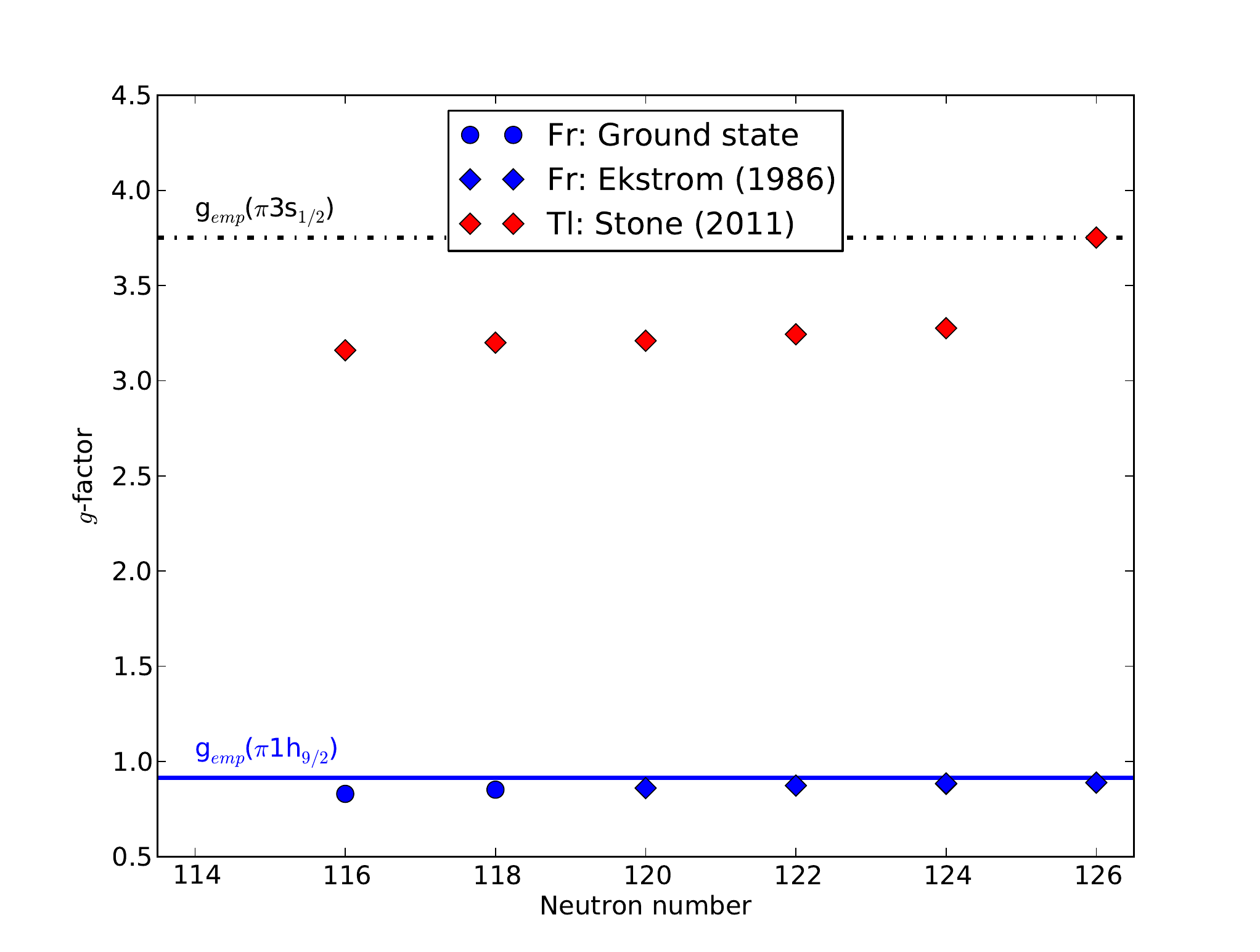}
\caption{\label{fig:CRIS_Fr_GF_Odd} $g$-factors for francium (blue)~\cite{Ekstrom1986, Gomez2008} and thallium (red) isotopes with odd-A~\cite{Stone2011}. The $g$-factors for the $\pi$3s$_{1/2}$ and $\pi$1h$_{9/2}$ proton orbitals have been calculated empirically. See text for details.}
\end{figure}

Figures~\ref{fig:CRIS_Fr_GF_Odd} and~\ref{fig:CRIS_Fr_GF_Even} show the experimental $g$-factors for odd-A and even-A francium isotopes, respectively. These plots present the CRIS data alongside the data from Ekstr\"om~\cite{Ekstrom1986}. The Ekstr\"om data has been re-evaluated with respect to the $\mu$($^{210}$Fr) measurement of Gomez~\cite{Gomez2008}.

In Fig.~\ref{fig:CRIS_Fr_GF_Odd}, the blue line represents the empirical $g$-factor ($g_{emp}$) of the odd-A isotopes for the single-particle occupation of the valence proton in the $\pi$1$h_{9/2}$ orbital. $g_{\text{emp}}$($\pi$1h$_{9/2}$) was determined from the magnetic moment of the single-particle state in $^{209}$Bi~\cite{Bastug1996}. Similarly, $g_{\text{emp}}$($\pi$3s$_{1/2}$) was estimated from the magnetic moment of the single-hole ground-state in $^{207}$Tl~\cite{Neugart1985}. From N~=~126 to 116, every isotope has a $g$-factor consistent with the proton occupying the $\pi$1h$_{9/2}$ orbital. This indicates that the 9/2$^-$ state remains the ground state, and the ($\pi$3s$^{-1}_{1/2}$)$_{1/2^+}$ proton intruder state has not yet inverted. This lowering in energy of the $\pi$3s$_{1/2}$ state to become the ground state would be apparent in the sudden increase in $g$-factor of the ground state, as illustrated by the black $g_{\text{emp}}$($\pi$3s$_{1/2}$) line.

\begin{figure}
\includegraphics[width = \columnwidth]{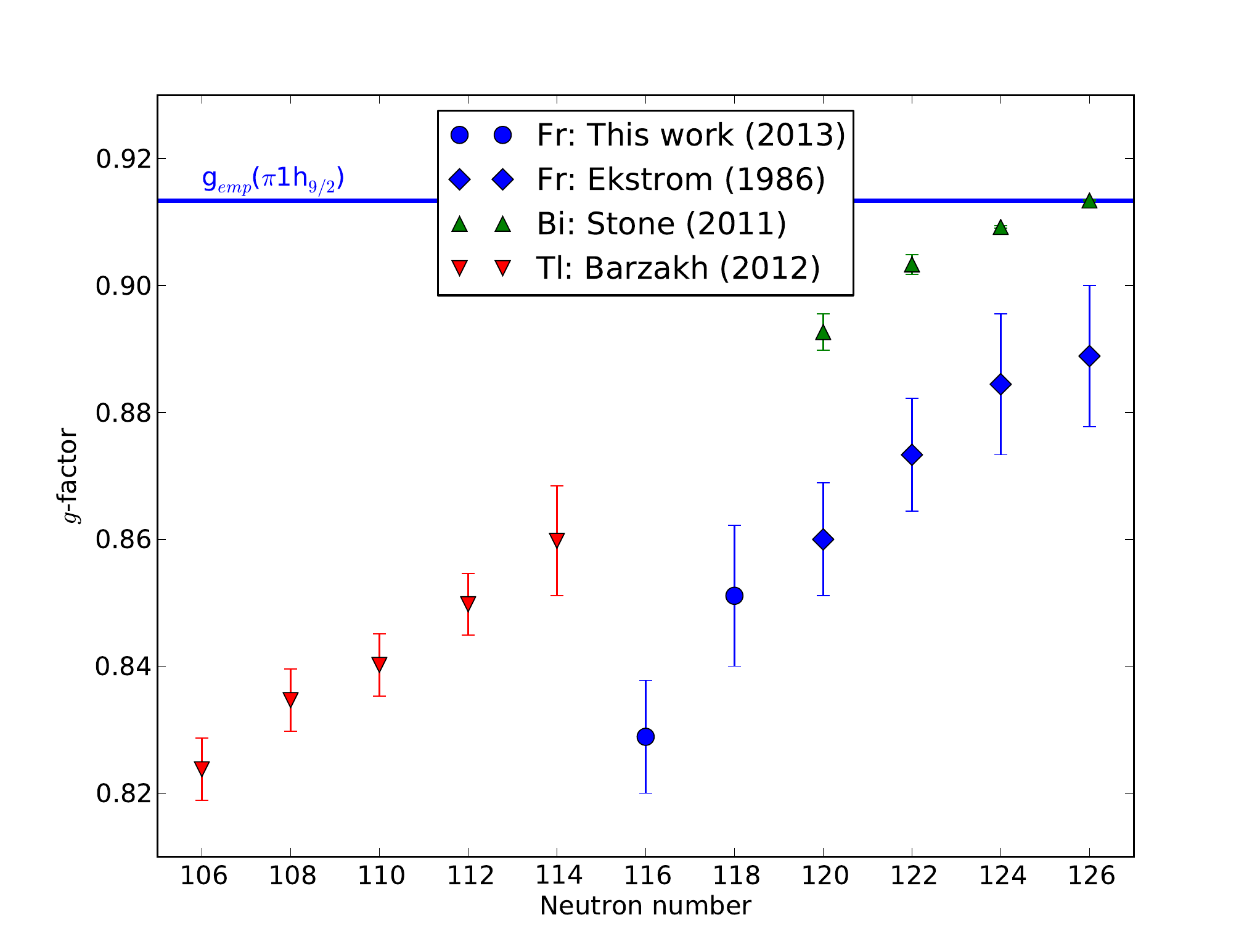}
\caption{\label{fig:CRIS_Fr_GF_Odd_zoom} Close up of the $g$-factors for francium (blue)~\cite{Ekstrom1986, Gomez2008}, bismuth (green)~\cite{Stone2011} and thallium (red) isotopes~\cite{Barzakh2012} with odd-A. The $g$-factor for the $\pi$1h$_{9/2}$ proton orbital has been calculated empirically. See text for details.}
\end{figure}

\begin{figure}
\includegraphics[width = \columnwidth]{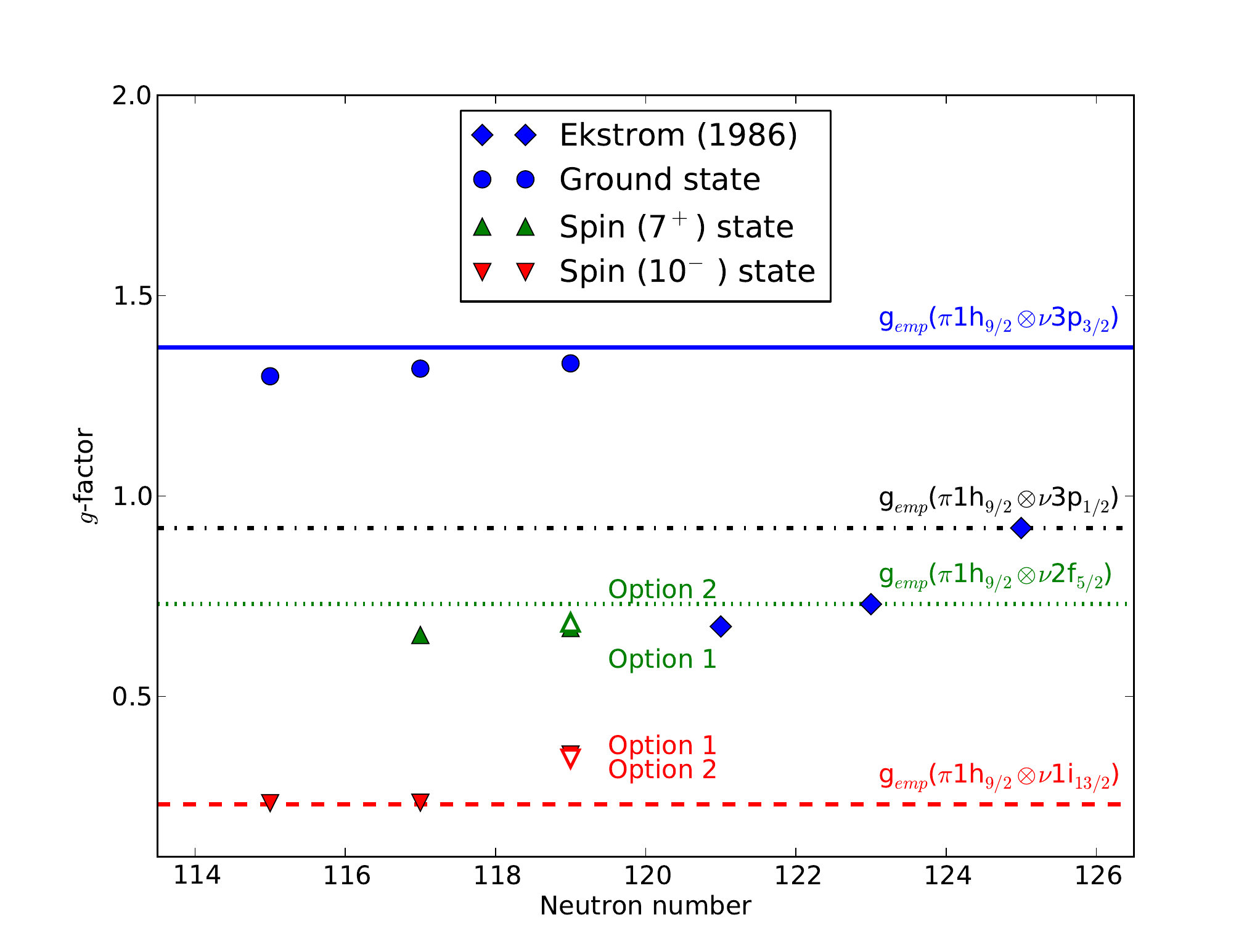}
\caption{\label{fig:CRIS_Fr_GF_Even} $g$-factors for francium isotopes with even-A~\cite{Ekstrom1986}: ground state (blue), spin (7$^+$) state (green) and spin (10$^-$) state (red). The $g$-factor for the coupling of the proton and neutron orbitals have been calculated empirically. See text for details.}
\end{figure}

Figure~\ref{fig:CRIS_Fr_GF_Odd} highlights the robustness of the Z~=~82 and N~=~126 shell closure with a shell-model description valid over a range of isotopes. A close-up of $g_{\text{emp}}$(1$\pi$1h$_{9/2}$) in Fig.~\ref{fig:CRIS_Fr_GF_Odd_zoom} illustrates that the $g$-factor is sensitive to bulk nuclear effects. The departure from the $g_{\text{emp}}$(1$\pi$1h$_{9/2}$) line shows the sensitivity of the $g$-factor to second-order core polarization in the odd-A thallium, bismuth and francium isotopes.  The systematic decrease in $g$-factor of francium is attributed to second-order core polarization associated with the presence of five valence particles, compared to one-particle (hole) in the bismuth (thallium) isotopes, enough to significantly weaken the shell closure. The linear trend observed in bismuth, thallium and francium (until N~=~118) is suggested to be related to the opening of the neutron shell, yet allowing for more neutron and proton-neutron correlations.

Further measurements towards the limit ./of stability are needed to better understand the prediction of the inversion of the $\pi$3s$_{1/2}$ intruder orbital with the $\pi$1h$_{9/2}$ ground state. A re-measurement of $^{203}$Fr could determine the presence of the spin 1/2$^+$ isomer (t$_{1/2}$ = 43(4)~ms~\cite{Jakobsson2013}), which was not observed during this experiment. 

The $g$-factors for the odd-odd francium isotopes are presented in Fig.~\ref{fig:CRIS_Fr_GF_Even}. With the coupling of the single valence proton in the $\pi$1$h_{9/2}$ orbital with a valence neutron, a large shell model space is available. The empirically calculated $g$-factors for the coupling of the $\pi$1h$_{9/2}$ proton with the valence neutrons are denoted by the colored lines. These $g$-factors were calculated from the additivity relation 
\begin{equation} g = \frac{1}{2} \Big [ g_p + g_n + (g_p - g_n) \frac{j_p(j_p +1) - j_n(j_n +1)}{I(I+1) } \Big ] ,\end{equation}
as outlined by Neyens~\cite{Neyens2003}. The empirical $g$-factors of the odd valence neutrons were calculated from the magnetic moments of neighbouring nuclei: $^{201}$Po for the blue $g_{\text{emp}}$($\pi$1h$_{9/2}\otimes\nu$3p$_{3/2}$) and red $g_{\text{emp}}$($\pi$1h$_{9/2}\otimes\nu$1i$_{13/2}$) line~\cite{Wouters1991}; $^{213}$Ra for the black $g_{\text{emp}}$($\pi$1h$_{9/2}\otimes\nu$3p$_{1/2}$) line; and $^{211}$Ra for the green $g_{\text{emp}}$($\pi$1h$_{9/2}\otimes\nu$2f$_{5/2}$) line~\cite{Ahmad1983}. The empirical $g$-factors for the valence proton in the $\pi$1h$_{9/2}$ orbital were calculated from the magnetic moment of the closest odd-A francium isotope ($^{203}$Fr and $^{213}$Fr respectively) from the CRIS data.

The ground state of $^{202,204,206}$Fr display similar $g$-factors, with the valence proton and neutron coupling to give a spin 3$^{(+)}$ state. The tentative configuration in literature of ($\pi$1h$_{9/2} \otimes \nu$2f$_{5/2}$)$_{3^+}$ for $^{202g}$Fr is based on the configuration of the (3$^+$) state in $^{194}$Bi (from favoured Fr-At-Bi alpha-decay chain systematics)~\cite{Zhu2008}. Similarly, the assignment of the same configuration for $^{204g}$Fr and $^{206g}$Fr is based on the alpha-decay systematics of neighbouring nuclei $^{196,198}$Bi. However, the initial assignment of $^{194g}$Bi was declared to be either ($\pi$1h$_{9/2} \otimes \nu$2f$_{5/2}$)$_{3^+}$ or ($\pi$1h$_{9/2} \otimes \nu$3p$_{3/2}$)$_{3^+}$~\cite{VanDuppen1991}. From the $g$-factors of the ground states of $^{202,204,206}$Fr, it is clear that the configuration of these states is indeed ($\pi$1h$_{9/2} \otimes \nu$3p$_{3/2}$)$_{3^+}$.

Figure~\ref{fig:CRIS_Fr_GF_Even} also presents the $g$-factors of $^{206m1}$Fr and $^{206m2}$Fr for option 1 and 2 (see Fig.~\ref{fig:CRIS_Fr206}). The first isomeric states of $^{204,206}$Fr (7$^+$) have a valence neutron that occupies the $\nu$2$f_{5/2}$ state. This coupling of the proton-particle neutron-hole results in a ($\pi$1h$_{9/2} \otimes \nu$2f$_{5/2}$)$_{7^+}$ configuration~\cite{Kondev2008}. For $^{202m}$Fr, $^{204m2}$Fr and $^{206m2}$Fr, the particle proton-neutron hole coupling result in a tentative ($\pi$1h$_{9/2} \otimes \nu$1i$_{13/2}$)$_{10^-}$ configuration assignment for each isomer~\cite{Chiara2010}. However, while the agreement of the $g$-factors of the spin (10$^-$) state in $^{202,204}$Fr point to a $\nu$1$i_{13/2}$ occupancy, the observed value for $^{206m2}$Fr is in disagreement with the $g$-factor of such a (10$^-$) state. The charge radius of $^{206m2}$Fr indicates a highly deformed configuration, where the single-particle description of the nucleus is no longer valid. This is consistent with the $g$-factor of this state: it is no longer obeying a simple shell-model description. This leads to the conclusion, that while a ($\pi$1h$_{9/2} \otimes \nu$1i$_{13/2}$)$_{10^-}$ configuration for $^{206m2}$Fr is suggested, the charge radii and magnetic moment point to a drastic change in the structure of this isomeric state. 

For completeness, the configurations of the odd-odd francium isotopes $^{208,210,212}$Fr are presented. The coupling of the valence proton and neutron in the $\pi$1h$_{9/2}$ and $\nu$2f$_{5/2}$ orbital in $^{208}$Fr and $^{210}$Fr leads to a ($\pi$1h$_{9/2} \otimes \nu$2f$_{5/2}$)$_{7^+}$ and ($\pi$1h$_{9/2} \otimes \nu$2f$_{5/2}$)$_{6^+}$  configuration respectively~\cite{Martin2007}. With the $\nu$2f$_{5/2}$ neutron orbital fully occupied, the valence neutron in $^{212}$Fr occupies the $\nu$3p$_{1/2}$ orbital, resulting in a ($\pi$1h$_{9/2} \otimes \nu$3p$_{1/2}$)$_{5^+}$ configuration~\cite{Browne2005}.

The agreement of the experimental and empirical $g$-factors, as shown in Figs.~\ref{fig:CRIS_Fr_GF_Odd}-\ref{fig:CRIS_Fr_GF_Even}, illustrates the suitability of the single-particle description of the neutron-deficient francium isotopes, with the exception of the (10$^-$) state in $^{206m2}$Fr. A model-independent spin and spectroscopic quadrupole moment determination is needed to clarify the nature of this isomeric state. The neutron-deficient francium isotopes display a single-particle nature where the additivity relation is still reliable.

\section{Conclusion and Outlook}

The hyperfine structures and isotope shifts of the neutron-deficient francium isotopes $^{202-206}$Fr with reference to $^{221}$Fr were measured with collinear resonance ionisation spectroscopy, and the change in mean-square charge radii and magnetic moments extracted. The selectivity of the alpha-decay patterns allowed the unambiguous identification of the hyperfine components of the low-lying isomers of $^{202,204}$Fr for the first time.

The resonant atomic transition of 7s~$^2$S$_{1/2} \rightarrow$ 8p~$^2$P$_{3/2}$ was probed, and the hyperfine $A_{S_{1/2}}$ factor measured. A King plot analysis of the 422.7-nm transition in francium allowed the atomic factors to be calibrated. The field and mass factors were determined to be F$_{422}$ = $-$20.670(210)~GHz/fm$^2$ and M$_{422}$ = +750(330)~GHz~amu, respectively.

The novel technique of decay-assisted laser spectroscopy in a collinear geometry was performed on the isotopes $^{202,204}$Fr. The decay spectroscopy station was utilized to identify the peaks in the hyperfine spectra of $^{202,204}$Fr. Alpha-tagging the hyperfine structure scan of $^{204}$Fr allowed the accurate determination of the nuclear observables of the three low-lying isomeric states and the determination of the branching ratios in the decay of $^{204m2}$Fr.

Analysis of the change in mean-square charge radii suggests an onset of collectivity that occurs at $^{203}$Fr (N~=~116). However, measurement of the spectroscopic quadrupole moment is required to determine the nature of the deformation (static or dynamic). The magnetic moments suggest that the single-particle description of the neutron-deficient francium isotopes still holds, with the exception of the (10$^-$) isomeric state of $^{206m2}$Fr. Based on the systematics of the region, the tentative assignment of the hyperfine structure peaks in $^{206}$Fr result in magnetic moments and mean-square charge radii that suggest a highly deformed state. Laser assisted nuclear decay spectroscopy of $^{206}$Fr would unambiguously determine their identity.

The occupation of the valence proton in the $\pi$1h$_{9/2}$ orbital has been suggested for all measured isotopes down to $^{202}$Fr, indicating the ($\pi$1s$_{1/2}^{-1}$)$_{1/2^+}$ intruder state does not yet invert with the $\pi$1h$_{9/2}$ orbital as the ground state. Further measurements of the very neutron-deficient francium isotopes towards $^{199}$Fr are required to fully determine the nature of the proton-intruder state. A laser linewidth of 1.5~GHz was enough to resolve the lower-state (7s~$^2$S$_{1/2}$) splitting of the hyperfine structure and measure the $A_{S_1/2}$ factor. In the future, the inclusion of a narrow-band laser system for the resonant-excitation step will enable the resolution of the upper-state (8p~$^2$P$_{3/2}$) splitting, providing the hyperfine $B_{P_{3/2}}$ factor. This will allow extraction of the spectroscopic quadrupole moment and determination of the nature of the deformation. 

Successful measurement of $^{202}$Fr was performed during this experiment, with a yield of 100 atoms per second. By pushing the limits of laser spectroscopy, further measurements of $^{201}$Fr (with a yield of 1 atom per second) and $^{200}$Fr (less than 1 atom per second) are thought to be possible. The ground state (9/2$^-$) of $^{201}$Fr has a half life of 53~ms and its isomer (1/2$^+$) a half life of 19~ms. By increasing the sensitivity of the CRIS technique, the presence of the 1/2$^+$ isomers in $^{201,203}$Fr can be confirmed. A positive identification will lead to nuclear-structure measurements that will determine (along with the verification of nuclear spin) the magnetic moments which are sensitive to the single-particle structure and thus to the ($\pi$3s$_{1/2}$)$_{1/2^+}$ proton intruder nature of these states. With sufficient resolution ($<$100~MHz), the spectroscopic quadrupole moment of these neutron-deficient states (with $I \ge 1$) will be directly measurable and the time-averaged static deformation can be determined.

The successful measurements performed by the CRIS experiment demonstrates the high sensitivity of the collinear resonance ionization technique. The decay spectroscopy station provides the ability to identify overlapping hyperfine structure and eventually perform laser assisted nuclear decay spectroscopy measurements on pure ground and isomeric-state beams~\cite{Lynch2012, Lynch2013}.

\section*{Acknowledgements}
The authors extend their thanks to the ISOLDE team for providing the beam, the GSI target lab for producing the carbon foils, and IKS-KU Leuven and The University of Manchester machine shops for their work. This work was supported by the IAP project P7/23 of the OSTC Belgium (BRIX network) and by the FWO-Vlaanderen (Belgium). The Manchester group was supported by the STFC consolidated grant ST/F012071/1 and continuation grant ST/J000159/1. K.T. Flanagan was supported by STFC Advanced Fellowship Scheme grant number ST/F012071/1. The authors would also like to thank Ed Schneiderman for continued support through donations to the Physics Department at NYU.

\bibliography{Bibliography}

\begin{thebibliography}{73}%
\makeatletter
\providecommand \@ifxundefined [1]{%
 \@ifx{#1\undefined}
}%
\providecommand \@ifnum [1]{%
 \ifnum #1\expandafter \@firstoftwo
 \else \expandafter \@secondoftwo
 \fi
}%
\providecommand \@ifx [1]{%
 \ifx #1\expandafter \@firstoftwo
 \else \expandafter \@secondoftwo
 \fi
}%
\providecommand \natexlab [1]{#1}%
\providecommand \enquote  [1]{``#1''}%
\providecommand \bibnamefont  [1]{#1}%
\providecommand \bibfnamefont [1]{#1}%
\providecommand \citenamefont [1]{#1}%
\providecommand \href@noop [0]{\@secondoftwo}%
\providecommand \href [0]{\begingroup \@sanitize@url \@href}%
\providecommand \@href[1]{\@@startlink{#1}\@@href}%
\providecommand \@@href[1]{\endgroup#1\@@endlink}%
\providecommand \@sanitize@url [0]{\catcode `\\12\catcode `\$12\catcode
  `\&12\catcode `\#12\catcode `\^12\catcode `\_12\catcode `\%12\relax}%
\providecommand \@@startlink[1]{}%
\providecommand \@@endlink[0]{}%
\providecommand \url  [0]{\begingroup\@sanitize@url \@url }%
\providecommand \@url [1]{\endgroup\@href {#1}{\urlprefix }}%
\providecommand \urlprefix  [0]{URL }%
\providecommand \Eprint [0]{\href }%
\providecommand \doibase [0]{http://dx.doi.org/}%
\providecommand \selectlanguage [0]{\@gobble}%
\providecommand \bibinfo  [0]{\@secondoftwo}%
\providecommand \bibfield  [0]{\@secondoftwo}%
\providecommand \translation [1]{[#1]}%
\providecommand \BibitemOpen [0]{}%
\providecommand \bibitemStop [0]{}%
\providecommand \bibitemNoStop [0]{.\EOS\space}%
\providecommand \EOS [0]{\spacefactor3000\relax}%
\providecommand \BibitemShut  [1]{\csname bibitem#1\endcsname}%
\let\auto@bib@innerbib\@empty
\bibitem [{\citenamefont {Cheal}\ and\ \citenamefont
  {Flanagan}(2010)}]{Cheal2010a}%
  \BibitemOpen
  \bibfield  {author} {\bibinfo {author} {\bibfnamefont {B.}~\bibnamefont
  {Cheal}}\ and\ \bibinfo {author} {\bibfnamefont {K.~T.}\ \bibnamefont
  {Flanagan}},\ }\href@noop {} {\bibfield  {journal} {\bibinfo  {journal} {J.
  Phys. G}\ }\textbf {\bibinfo {volume} {37}},\ \bibinfo {pages} {113101}
  (\bibinfo {year} {2010})}\BibitemShut {NoStop}%
\bibitem [{\citenamefont {Blaum}\ \emph {et~al.}(2013)\citenamefont {Blaum},
  \citenamefont {Dilling},\ and\ \citenamefont
  {N\"ortersh\"auser}}]{Blaum2013}%
  \BibitemOpen
  \bibfield  {author} {\bibinfo {author} {\bibfnamefont {K.}~\bibnamefont
  {Blaum}}, \bibinfo {author} {\bibfnamefont {J.}~\bibnamefont {Dilling}}, \
  and\ \bibinfo {author} {\bibfnamefont {W.}~\bibnamefont
  {N\"ortersh\"auser}},\ }\href
  {http://stacks.iop.org/1402-4896/2013/i=T152/a=014017} {\bibfield  {journal}
  {\bibinfo  {journal} {Physica Scripta}\ }\textbf {\bibinfo {volume} {2013}},\
  \bibinfo {pages} {014017} (\bibinfo {year} {2013})}\BibitemShut {NoStop}%
\bibitem [{\citenamefont {Procter}\ and\ \citenamefont
  {Flanagan}(2013)}]{Procter2013}%
  \BibitemOpen
  \bibfield  {author} {\bibinfo {author} {\bibfnamefont {T.~J.}\ \bibnamefont
  {Procter}}\ and\ \bibinfo {author} {\bibfnamefont {K.~T.}\ \bibnamefont
  {Flanagan}},\ }\href {http://dx.doi.org/10.1007/s10751-013-0833-6} {\bibfield
   {journal} {\bibinfo  {journal} {Hyperfine Interact.}\ }\textbf {\bibinfo
  {volume} {216}},\ \bibinfo {pages} {89} (\bibinfo {year} {2013})}\BibitemShut
  {NoStop}%
\bibitem [{\citenamefont {Liberman}\ \emph {et~al.}(1978)\citenamefont
  {Liberman} \emph {et~al.}}]{Liberman1978}%
  \BibitemOpen
  \bibfield  {author} {\bibinfo {author} {\bibfnamefont {S.}~\bibnamefont
  {Liberman}} \emph {et~al.},\ }\href@noop {} {\bibfield  {journal} {\bibinfo
  {journal} {C. R. Acad. Sci. Paris Ser. B}\ }\textbf {\bibinfo {volume}
  {286}},\ \bibinfo {pages} {353} (\bibinfo {year} {1978})}\BibitemShut
  {NoStop}%
\bibitem [{\citenamefont {Liberman}\ \emph {et~al.}(1980)\citenamefont
  {Liberman} \emph {et~al.}}]{Liberman1980}%
  \BibitemOpen
  \bibfield  {author} {\bibinfo {author} {\bibfnamefont {S.}~\bibnamefont
  {Liberman}} \emph {et~al.},\ }\href@noop {} {\bibfield  {journal} {\bibinfo
  {journal} {Phys. Rev. A}\ }\textbf {\bibinfo {volume} {22}},\ \bibinfo
  {pages} {2732} (\bibinfo {year} {1980})}\BibitemShut {NoStop}%
\bibitem [{\citenamefont {Yagoda}(1932)}]{Yagoda1932}%
  \BibitemOpen
  \bibfield  {author} {\bibinfo {author} {\bibfnamefont {H.}~\bibnamefont
  {Yagoda}},\ }\href@noop {} {\bibfield  {journal} {\bibinfo  {journal} {Phys.
  Rev.}\ }\textbf {\bibinfo {volume} {40}},\ \bibinfo {pages} {1017} (\bibinfo
  {year} {1932})}\BibitemShut {NoStop}%
\bibitem [{\citenamefont {Coc}\ \emph {et~al.}(1985)\citenamefont {Coc} \emph
  {et~al.}}]{Coc1985}%
  \BibitemOpen
  \bibfield  {author} {\bibinfo {author} {\bibfnamefont {A.}~\bibnamefont
  {Coc}} \emph {et~al.},\ }\href@noop {} {\bibfield  {journal} {\bibinfo
  {journal} {Phys. Lett. B}\ }\textbf {\bibinfo {volume} {163}},\ \bibinfo
  {pages} {66 } (\bibinfo {year} {1985})}\BibitemShut {NoStop}%
\bibitem [{\citenamefont {Touchard}\ \emph {et~al.}(1984)\citenamefont
  {Touchard} \emph {et~al.}}]{Touchard1984}%
  \BibitemOpen
  \bibfield  {author} {\bibinfo {author} {\bibfnamefont {F.}~\bibnamefont
  {Touchard}} \emph {et~al.},\ }\href@noop {} {\bibfield  {journal} {\bibinfo
  {journal} {Atomic Masses and Fundamental Constants 7}\ ,\ \bibinfo {pages}
  {353}} (\bibinfo {year} {1984})}\BibitemShut {NoStop}%
\bibitem [{\citenamefont {Bauche}\ \emph {et~al.}(1986)\citenamefont {Bauche}
  \emph {et~al.}}]{Bauche1986}%
  \BibitemOpen
  \bibfield  {author} {\bibinfo {author} {\bibfnamefont {J.}~\bibnamefont
  {Bauche}} \emph {et~al.},\ }\href@noop {} {\bibfield  {journal} {\bibinfo
  {journal} {J. Phys. B: At. Mol. Opt. Phys.}\ }\textbf {\bibinfo {volume}
  {19}},\ \bibinfo {pages} {L593} (\bibinfo {year} {1986})}\BibitemShut
  {NoStop}%
\bibitem [{\citenamefont {Duong}\ \emph {et~al.}(1987)\citenamefont {Duong}
  \emph {et~al.}}]{Duong1987}%
  \BibitemOpen
  \bibfield  {author} {\bibinfo {author} {\bibfnamefont {H.~T.}\ \bibnamefont
  {Duong}} \emph {et~al.},\ }\href@noop {} {\bibfield  {journal} {\bibinfo
  {journal} {Europhys. Lett.}\ }\textbf {\bibinfo {volume} {3}},\ \bibinfo
  {pages} {175} (\bibinfo {year} {1987})}\BibitemShut {NoStop}%
\bibitem [{\citenamefont {Andreev}\ \emph {et~al.}(1987)\citenamefont
  {Andreev}, \citenamefont {Mishin},\ and\ \citenamefont
  {Letokhov}}]{Andreev1987}%
  \BibitemOpen
  \bibfield  {author} {\bibinfo {author} {\bibfnamefont {S.~V.}\ \bibnamefont
  {Andreev}}, \bibinfo {author} {\bibfnamefont {V.~I.}\ \bibnamefont {Mishin}},
  \ and\ \bibinfo {author} {\bibfnamefont {V.~S.}\ \bibnamefont {Letokhov}},\
  }\href@noop {} {\bibfield  {journal} {\bibinfo  {journal} {Phys. Rev. Lett.}\
  }\textbf {\bibinfo {volume} {59}},\ \bibinfo {pages} {1274} (\bibinfo {year}
  {1987})}\BibitemShut {NoStop}%
\bibitem [{\citenamefont {Kudriavtsev}\ and\ \citenamefont
  {Letokhov}(1982)}]{Kudriavtsev1982}%
  \BibitemOpen
  \bibfield  {author} {\bibinfo {author} {\bibfnamefont {Y.}~\bibnamefont
  {Kudriavtsev}}\ and\ \bibinfo {author} {\bibfnamefont {V.}~\bibnamefont
  {Letokhov}},\ }\href {\doibase 10.1007/BF00688671} {\bibfield  {journal}
  {\bibinfo  {journal} {Applied Physics B}\ }\textbf {\bibinfo {volume} {29}},\
  \bibinfo {pages} {219} (\bibinfo {year} {1982})}\BibitemShut {NoStop}%
\bibitem [{\citenamefont {Schulz}\ \emph {et~al.}(1991)\citenamefont {Schulz}
  \emph {et~al.}}]{Schulz1991}%
  \BibitemOpen
  \bibfield  {author} {\bibinfo {author} {\bibfnamefont {C.}~\bibnamefont
  {Schulz}} \emph {et~al.},\ }\href
  {http://stacks.iop.org/0953-4075/24/i=22/a=020} {\bibfield  {journal}
  {\bibinfo  {journal} {J. Phys. B: At. Mol. Opt. Phys.}\ }\textbf {\bibinfo
  {volume} {24}},\ \bibinfo {pages} {4831} (\bibinfo {year}
  {1991})}\BibitemShut {NoStop}%
\bibitem [{\citenamefont {Davids}\ \emph {et~al.}(1996)\citenamefont {Davids}
  \emph {et~al.}}]{Davids1996}%
  \BibitemOpen
  \bibfield  {author} {\bibinfo {author} {\bibfnamefont {C.~N.}\ \bibnamefont
  {Davids}} \emph {et~al.},\ }\href {\doibase 10.1103/PhysRevLett.76.592}
  {\bibfield  {journal} {\bibinfo  {journal} {Phys. Rev. Lett.}\ }\textbf
  {\bibinfo {volume} {76}},\ \bibinfo {pages} {592} (\bibinfo {year}
  {1996})}\BibitemShut {NoStop}%
\bibitem [{\citenamefont {Kettunen}\ \emph {et~al.}(2003)\citenamefont
  {Kettunen} \emph {et~al.}}]{Kettunen2003}%
  \BibitemOpen
  \bibfield  {author} {\bibinfo {author} {\bibfnamefont {H.}~\bibnamefont
  {Kettunen}} \emph {et~al.},\ }\href@noop {} {\bibfield  {journal} {\bibinfo
  {journal} {Eur. Phys. J. A}\ }\textbf {\bibinfo {volume} {16}},\ \bibinfo
  {pages} {457} (\bibinfo {year} {2003})}\BibitemShut {NoStop}%
\bibitem [{\citenamefont {Uusitalo}\ \emph {et~al.}(2005)\citenamefont
  {Uusitalo} \emph {et~al.}}]{Uusitalo2005}%
  \BibitemOpen
  \bibfield  {author} {\bibinfo {author} {\bibfnamefont {J.}~\bibnamefont
  {Uusitalo}} \emph {et~al.},\ }\href {\doibase 10.1103/PhysRevC.71.024306}
  {\bibfield  {journal} {\bibinfo  {journal} {Phys. Rev. C}\ }\textbf {\bibinfo
  {volume} {71}},\ \bibinfo {pages} {024306} (\bibinfo {year}
  {2005})}\BibitemShut {NoStop}%
\bibitem [{\citenamefont {Jakobsson}\ \emph {et~al.}(2012)\citenamefont
  {Jakobsson} \emph {et~al.}}]{Jakobsson2012}%
  \BibitemOpen
  \bibfield  {author} {\bibinfo {author} {\bibfnamefont {U.}~\bibnamefont
  {Jakobsson}} \emph {et~al.},\ }\href {\doibase 10.1103/PhysRevC.85.014309}
  {\bibfield  {journal} {\bibinfo  {journal} {Phys. Rev. C}\ }\textbf {\bibinfo
  {volume} {85}},\ \bibinfo {pages} {014309} (\bibinfo {year}
  {2012})}\BibitemShut {NoStop}%
\bibitem [{\citenamefont {Huyse}\ \emph {et~al.}(1992)\citenamefont {Huyse},
  \citenamefont {Decrock}, \citenamefont {Dendooven}, \citenamefont {Reusen},
  \citenamefont {Van~Duppen},\ and\ \citenamefont {Wauters}}]{Huyse1992}%
  \BibitemOpen
  \bibfield  {author} {\bibinfo {author} {\bibfnamefont {M.}~\bibnamefont
  {Huyse}}, \bibinfo {author} {\bibfnamefont {P.}~\bibnamefont {Decrock}},
  \bibinfo {author} {\bibfnamefont {P.}~\bibnamefont {Dendooven}}, \bibinfo
  {author} {\bibfnamefont {G.}~\bibnamefont {Reusen}}, \bibinfo {author}
  {\bibfnamefont {P.}~\bibnamefont {Van~Duppen}}, \ and\ \bibinfo {author}
  {\bibfnamefont {J.}~\bibnamefont {Wauters}},\ }\href {\doibase
  10.1103/PhysRevC.46.1209} {\bibfield  {journal} {\bibinfo  {journal} {Phys.
  Rev. C}\ }\textbf {\bibinfo {volume} {46}},\ \bibinfo {pages} {1209}
  (\bibinfo {year} {1992})}\BibitemShut {NoStop}%
\bibitem [{\citenamefont {Jakobsson}\ \emph {et~al.}(2013)\citenamefont
  {Jakobsson} \emph {et~al.}}]{Jakobsson2013}%
  \BibitemOpen
  \bibfield  {author} {\bibinfo {author} {\bibfnamefont {U.}~\bibnamefont
  {Jakobsson}} \emph {et~al.},\ }\href {\doibase 10.1103/PhysRevC.87.054320}
  {\bibfield  {journal} {\bibinfo  {journal} {Phys. Rev. C}\ }\textbf {\bibinfo
  {volume} {87}},\ \bibinfo {pages} {054320} (\bibinfo {year}
  {2013})}\BibitemShut {NoStop}%
\bibitem [{\citenamefont {Voss}\ \emph {et~al.}(2013)\citenamefont {Voss},
  \citenamefont {Pearson}, \citenamefont {Billowes}, \citenamefont {Buchinger},
  \citenamefont {Cheal}, \citenamefont {Crawford}, \citenamefont {Kwiatkowski},
  \citenamefont {Levy},\ and\ \citenamefont {Shelbaya}}]{Voss2013}%
  \BibitemOpen
  \bibfield  {author} {\bibinfo {author} {\bibfnamefont {A.}~\bibnamefont
  {Voss}}, \bibinfo {author} {\bibfnamefont {M.~R.}\ \bibnamefont {Pearson}},
  \bibinfo {author} {\bibfnamefont {J.}~\bibnamefont {Billowes}}, \bibinfo
  {author} {\bibfnamefont {F.}~\bibnamefont {Buchinger}}, \bibinfo {author}
  {\bibfnamefont {B.}~\bibnamefont {Cheal}}, \bibinfo {author} {\bibfnamefont
  {J.~E.}\ \bibnamefont {Crawford}}, \bibinfo {author} {\bibfnamefont {A.~A.}\
  \bibnamefont {Kwiatkowski}}, \bibinfo {author} {\bibfnamefont {C.~D.~P.}\
  \bibnamefont {Levy}}, \ and\ \bibinfo {author} {\bibfnamefont
  {O.}~\bibnamefont {Shelbaya}},\ }\href {\doibase
  10.1103/PhysRevLett.111.122501} {\bibfield  {journal} {\bibinfo  {journal}
  {Phys. Rev. Lett.}\ }\textbf {\bibinfo {volume} {111}},\ \bibinfo {pages}
  {122501} (\bibinfo {year} {2013})}\BibitemShut {NoStop}%
\bibitem [{\citenamefont {Fedosseev}\ \emph
  {et~al.}(2012{\natexlab{a}})\citenamefont {Fedosseev}, \citenamefont
  {Kudryavtsev},\ and\ \citenamefont {Mishin}}]{Fedosseev2012a}%
  \BibitemOpen
  \bibfield  {author} {\bibinfo {author} {\bibfnamefont {V.~N.}\ \bibnamefont
  {Fedosseev}}, \bibinfo {author} {\bibfnamefont {Y.}~\bibnamefont
  {Kudryavtsev}}, \ and\ \bibinfo {author} {\bibfnamefont {V.~I.}\ \bibnamefont
  {Mishin}},\ }\href {http://stacks.iop.org/1402-4896/85/i=5/a=058104}
  {\bibfield  {journal} {\bibinfo  {journal} {Phys. Scripta}\ }\textbf
  {\bibinfo {volume} {85}},\ \bibinfo {pages} {058104} (\bibinfo {year}
  {2012}{\natexlab{a}})}\BibitemShut {NoStop}%
\bibitem [{\citenamefont {Weissman}\ \emph {et~al.}(2002)\citenamefont
  {Weissman} \emph {et~al.}}]{Weissmann2002}%
  \BibitemOpen
  \bibfield  {author} {\bibinfo {author} {\bibfnamefont {L.}~\bibnamefont
  {Weissman}} \emph {et~al.},\ }\href {\doibase 10.1103/PhysRevC.65.024315}
  {\bibfield  {journal} {\bibinfo  {journal} {Phys. Rev. C}\ }\textbf {\bibinfo
  {volume} {65}},\ \bibinfo {pages} {024315} (\bibinfo {year}
  {2002})}\BibitemShut {NoStop}%
\bibitem [{\citenamefont {Stefanescu}\ \emph {et~al.}(2007)\citenamefont
  {Stefanescu} \emph {et~al.}}]{Stefanescu2007}%
  \BibitemOpen
  \bibfield  {author} {\bibinfo {author} {\bibfnamefont {I.}~\bibnamefont
  {Stefanescu}} \emph {et~al.},\ }\href {\doibase
  10.1103/PhysRevLett.98.122701} {\bibfield  {journal} {\bibinfo  {journal}
  {Phys. Rev. Lett.}\ }\textbf {\bibinfo {volume} {98}},\ \bibinfo {pages}
  {122701} (\bibinfo {year} {2007})}\BibitemShut {NoStop}%
\bibitem [{\citenamefont {Van~Roosbroeck}\ \emph {et~al.}(2004)\citenamefont
  {Van~Roosbroeck} \emph {et~al.}}]{VanRoosbroeck2004}%
  \BibitemOpen
  \bibfield  {author} {\bibinfo {author} {\bibfnamefont {J.}~\bibnamefont
  {Van~Roosbroeck}} \emph {et~al.},\ }\href {\doibase
  10.1103/PhysRevLett.92.112501} {\bibfield  {journal} {\bibinfo  {journal}
  {Phys. Rev. Lett.}\ }\textbf {\bibinfo {volume} {92}},\ \bibinfo {pages}
  {112501} (\bibinfo {year} {2004})}\BibitemShut {NoStop}%
\bibitem [{\citenamefont {Kugler}(2000)}]{Kugler2000}%
  \BibitemOpen
  \bibfield  {author} {\bibinfo {author} {\bibfnamefont {E.}~\bibnamefont
  {Kugler}},\ }\href@noop {} {\bibfield  {journal} {\bibinfo  {journal}
  {Hyperfine Interact.}\ }\textbf {\bibinfo {volume} {129}},\ \bibinfo {pages}
  {23} (\bibinfo {year} {2000})}\BibitemShut {NoStop}%
\bibitem [{\citenamefont {Jokinen}\ \emph {et~al.}(2003)\citenamefont
  {Jokinen}, \citenamefont {Lindroos}, \citenamefont {Molin},\ and\
  \citenamefont {Petersson}}]{Jokinen2003}%
  \BibitemOpen
  \bibfield  {author} {\bibinfo {author} {\bibfnamefont {A.}~\bibnamefont
  {Jokinen}}, \bibinfo {author} {\bibfnamefont {M.}~\bibnamefont {Lindroos}},
  \bibinfo {author} {\bibfnamefont {E.}~\bibnamefont {Molin}}, \ and\ \bibinfo
  {author} {\bibfnamefont {M.}~\bibnamefont {Petersson}},\ }\href@noop {}
  {\bibfield  {journal} {\bibinfo  {journal} {Nucl. Instrum. Methods Phys. Res.
  B}\ }\textbf {\bibinfo {volume} {204}},\ \bibinfo {pages} {86 } (\bibinfo
  {year} {2003})}\BibitemShut {NoStop}%
\bibitem [{\citenamefont {Man\'e}\ \emph {et~al.}(2009)\citenamefont {Man\'e}
  \emph {et~al.}}]{Mane2009}%
  \BibitemOpen
  \bibfield  {author} {\bibinfo {author} {\bibfnamefont {E.}~\bibnamefont
  {Man\'e}} \emph {et~al.},\ }\href@noop {} {\bibfield  {journal} {\bibinfo
  {journal} {Eur. Phys. J. A}\ }\textbf {\bibinfo {volume} {42}},\ \bibinfo
  {pages} {503} (\bibinfo {year} {2009})}\BibitemShut {NoStop}%
\bibitem [{\citenamefont {Haynes}(2013)}]{Handbook2013}%
  \BibitemOpen
  \bibfield  {author} {\bibinfo {author} {\bibfnamefont {W.~H.}\ \bibnamefont
  {Haynes}},\ }\href@noop {} {\emph {\bibinfo {title} {Handbook of Chemistry
  and Physics}}},\ \bibinfo {edition} {94th}\ ed.\ (\bibinfo  {publisher}
  {CRC},\ \bibinfo {year} {2013})\BibitemShut {NoStop}%
\bibitem [{\citenamefont {Fedosseev}\ \emph
  {et~al.}(2012{\natexlab{b}})\citenamefont {Fedosseev} \emph
  {et~al.}}]{Fedosseev2012b}%
  \BibitemOpen
  \bibfield  {author} {\bibinfo {author} {\bibfnamefont {V.~N.}\ \bibnamefont
  {Fedosseev}} \emph {et~al.},\ }\href@noop {} {\bibfield  {journal} {\bibinfo
  {journal} {Rev. Sci. Instrum.}\ }\textbf {\bibinfo {volume} {83}},\ \bibinfo
  {eid} {02A903} (\bibinfo {year} {2012}{\natexlab{b}})}\BibitemShut {NoStop}%
\bibitem [{\citenamefont {Rothe}\ \emph {et~al.}(2013)\citenamefont {Rothe},
  \citenamefont {Fedosseev}, \citenamefont {Kron}, \citenamefont {Marsh},
  \citenamefont {Rossel},\ and\ \citenamefont {Wendt}}]{Rothe2013}%
  \BibitemOpen
  \bibfield  {author} {\bibinfo {author} {\bibfnamefont {S.}~\bibnamefont
  {Rothe}}, \bibinfo {author} {\bibfnamefont {V.~N.}\ \bibnamefont
  {Fedosseev}}, \bibinfo {author} {\bibfnamefont {T.}~\bibnamefont {Kron}},
  \bibinfo {author} {\bibfnamefont {B.~A.}\ \bibnamefont {Marsh}}, \bibinfo
  {author} {\bibfnamefont {R.~E.}\ \bibnamefont {Rossel}}, \ and\ \bibinfo
  {author} {\bibfnamefont {K.~D.~A.}\ \bibnamefont {Wendt}},\ }\href@noop {}
  {\bibfield  {journal} {\bibinfo  {journal} {Nucl. Instrum. Methods Phys. Res.
  B}\ }\textbf {\bibinfo {volume} {317, Part B}},\ \bibinfo {pages} {561 }
  (\bibinfo {year} {2013})}\BibitemShut {NoStop}%
\bibitem [{\citenamefont {Rossel}\ \emph {et~al.}(2013)\citenamefont {Rossel},
  \citenamefont {Fedosseev}, \citenamefont {Marsh}, \citenamefont {Richter},
  \citenamefont {Rothe},\ and\ \citenamefont {Wendt}}]{Rossel2013}%
  \BibitemOpen
  \bibfield  {author} {\bibinfo {author} {\bibfnamefont {R.~E.}\ \bibnamefont
  {Rossel}}, \bibinfo {author} {\bibfnamefont {V.~N.}\ \bibnamefont
  {Fedosseev}}, \bibinfo {author} {\bibfnamefont {B.~A.}\ \bibnamefont
  {Marsh}}, \bibinfo {author} {\bibfnamefont {D.}~\bibnamefont {Richter}},
  \bibinfo {author} {\bibfnamefont {S.}~\bibnamefont {Rothe}}, \ and\ \bibinfo
  {author} {\bibfnamefont {K.~D.~A.}\ \bibnamefont {Wendt}},\ }\href@noop {}
  {\bibfield  {journal} {\bibinfo  {journal} {Nucl. Instrum. Methods Phys. Res.
  B}\ }\textbf {\bibinfo {volume} {317, Part B}},\ \bibinfo {pages} {557 }
  (\bibinfo {year} {2013})}\BibitemShut {NoStop}%
\bibitem [{\citenamefont {Lynch}\ \emph
  {et~al.}(2013{\natexlab{a}})\citenamefont {Lynch}, \citenamefont {Cocolios},\
  and\ \citenamefont {Rajabali}}]{Lynch2013}%
  \BibitemOpen
  \bibfield  {author} {\bibinfo {author} {\bibfnamefont {K.~M.}\ \bibnamefont
  {Lynch}}, \bibinfo {author} {\bibfnamefont {T.~E.}\ \bibnamefont {Cocolios}},
  \ and\ \bibinfo {author} {\bibfnamefont {M.~M.}\ \bibnamefont {Rajabali}},\
  }\href@noop {} {\bibfield  {journal} {\bibinfo  {journal} {Hyperfine
  Interact.}\ }\textbf {\bibinfo {volume} {216}},\ \bibinfo {pages} {95}
  (\bibinfo {year} {2013}{\natexlab{a}})}\BibitemShut {NoStop}%
\bibitem [{\citenamefont {Rajabali}\ \emph {et~al.}(2013)\citenamefont
  {Rajabali} \emph {et~al.}}]{Rajabali2013}%
  \BibitemOpen
  \bibfield  {author} {\bibinfo {author} {\bibfnamefont {M.~M.}\ \bibnamefont
  {Rajabali}} \emph {et~al.},\ }\href@noop {} {\bibfield  {journal} {\bibinfo
  {journal} {Nucl. Instrum. Methods Phys. Res. A}\ }\textbf {\bibinfo {volume}
  {707}},\ \bibinfo {pages} {35 } (\bibinfo {year} {2013})}\BibitemShut
  {NoStop}%
\bibitem [{\citenamefont {Dendooven}(1992)}]{DendoovenThesis}%
  \BibitemOpen
  \bibfield  {author} {\bibinfo {author} {\bibfnamefont {P.}~\bibnamefont
  {Dendooven}},\ }\href@noop {} {Ph.D. thesis},\ \bibinfo  {school} {{IKS}, KU
  Leuven} (\bibinfo {year} {1992})\BibitemShut {NoStop}%
\bibitem [{\citenamefont {Andreyev}\ \emph {et~al.}(2010)\citenamefont
  {Andreyev} \emph {et~al.}}]{Andreyev2010}%
  \BibitemOpen
  \bibfield  {author} {\bibinfo {author} {\bibfnamefont {A.~N.}\ \bibnamefont
  {Andreyev}} \emph {et~al.},\ }\href@noop {} {\bibfield  {journal} {\bibinfo
  {journal} {Phys. Rev. Lett.}\ }\textbf {\bibinfo {volume} {105}},\ \bibinfo
  {pages} {252502} (\bibinfo {year} {2010})}\BibitemShut {NoStop}%
\bibitem [{\citenamefont {Elseviers}\ \emph {et~al.}(2013)\citenamefont
  {Elseviers} \emph {et~al.}}]{Elseviers2013}%
  \BibitemOpen
  \bibfield  {author} {\bibinfo {author} {\bibfnamefont {J.}~\bibnamefont
  {Elseviers}} \emph {et~al.},\ }\href@noop {} {\bibfield  {journal} {\bibinfo
  {journal} {Phys. Rev. C}\ }\textbf {\bibinfo {volume} {88}},\ \bibinfo
  {pages} {044321} (\bibinfo {year} {2013})}\BibitemShut {NoStop}%
\bibitem [{\citenamefont {Lommel}\ \emph {et~al.}(2002)\citenamefont {Lommel},
  \citenamefont {Hartmann}, \citenamefont {Kindler}, \citenamefont {Klemm},\
  and\ \citenamefont {Steiner}}]{Lommel2002}%
  \BibitemOpen
  \bibfield  {author} {\bibinfo {author} {\bibfnamefont {B.}~\bibnamefont
  {Lommel}}, \bibinfo {author} {\bibfnamefont {W.}~\bibnamefont {Hartmann}},
  \bibinfo {author} {\bibfnamefont {B.}~\bibnamefont {Kindler}}, \bibinfo
  {author} {\bibfnamefont {J.}~\bibnamefont {Klemm}}, \ and\ \bibinfo {author}
  {\bibfnamefont {J.}~\bibnamefont {Steiner}},\ }\href@noop {} {\bibfield
  {journal} {\bibinfo  {journal} {Nucl. Instrum. Methods Phys. Res. A}\
  }\textbf {\bibinfo {volume} {480}},\ \bibinfo {pages} {199 } (\bibinfo {year}
  {2002})}\BibitemShut {NoStop}%
\bibitem [{\citenamefont {Hennig}\ \emph {et~al.}(2007)\citenamefont {Hennig},
  \citenamefont {Tan}, \citenamefont {Walby}, \citenamefont {Grudberg},
  \citenamefont {Fallu-Labruyere}, \citenamefont {Warburton}, \citenamefont
  {Vaman}, \citenamefont {Starosta},\ and\ \citenamefont
  {Miller}}]{Hennig2007}%
  \BibitemOpen
  \bibfield  {author} {\bibinfo {author} {\bibfnamefont {W.}~\bibnamefont
  {Hennig}}, \bibinfo {author} {\bibfnamefont {H.}~\bibnamefont {Tan}},
  \bibinfo {author} {\bibfnamefont {M.}~\bibnamefont {Walby}}, \bibinfo
  {author} {\bibfnamefont {P.}~\bibnamefont {Grudberg}}, \bibinfo {author}
  {\bibfnamefont {A.}~\bibnamefont {Fallu-Labruyere}}, \bibinfo {author}
  {\bibfnamefont {W.~K.}\ \bibnamefont {Warburton}}, \bibinfo {author}
  {\bibfnamefont {C.}~\bibnamefont {Vaman}}, \bibinfo {author} {\bibfnamefont
  {K.}~\bibnamefont {Starosta}}, \ and\ \bibinfo {author} {\bibfnamefont
  {D.}~\bibnamefont {Miller}},\ }\href@noop {} {\bibfield  {journal} {\bibinfo
  {journal} {Nucl. Instrum. Methods Phys. Res. B}\ }\textbf {\bibinfo {volume}
  {261}},\ \bibinfo {pages} {1000 } (\bibinfo {year} {2007})}\BibitemShut
  {NoStop}%
\bibitem [{\citenamefont {Flanagan}\ \emph {et~al.}(2013)\citenamefont
  {Flanagan} \emph {et~al.}}]{Flanagan2013}%
  \BibitemOpen
  \bibfield  {author} {\bibinfo {author} {\bibfnamefont {K.~T.}\ \bibnamefont
  {Flanagan}} \emph {et~al.},\ }\href@noop {} {\bibfield  {journal} {\bibinfo
  {journal} {Phys. Rev. Lett.}\ }\textbf {\bibinfo {volume} {111}},\ \bibinfo
  {pages} {212501} (\bibinfo {year} {2013})}\BibitemShut {NoStop}%
\bibitem [{\citenamefont {Budin\u{c}evi\'c}\ \emph {et~al.}()\citenamefont
  {Budin\u{c}evi\'c} \emph {et~al.}}]{Budincevic2014}%
  \BibitemOpen
  \bibfield  {author} {\bibinfo {author} {\bibfnamefont {I.}~\bibnamefont
  {Budin\u{c}evi\'c}} \emph {et~al.},\ }\href@noop {} {}\bibinfo {note}
  {(unpublished)}\BibitemShut {NoStop}%
\bibitem [{\citenamefont {Cocolios}\ \emph {et~al.}(2013)\citenamefont
  {Cocolios} \emph {et~al.}}]{Cocolios2013b}%
  \BibitemOpen
  \bibfield  {author} {\bibinfo {author} {\bibfnamefont {T.~E.}\ \bibnamefont
  {Cocolios}} \emph {et~al.},\ }\href@noop {} {\bibfield  {journal} {\bibinfo
  {journal} {Nucl. Instrum. Methods Phys. Res. B}\ }\textbf {\bibinfo {volume}
  {317, Part B}},\ \bibinfo {pages} {565 } (\bibinfo {year}
  {2013})}\BibitemShut {NoStop}%
\bibitem [{\citenamefont {Lynch}(2013)}]{MyThesis}%
  \BibitemOpen
  \bibfield  {author} {\bibinfo {author} {\bibfnamefont {K.~M.}\ \bibnamefont
  {Lynch}},\ }\href@noop {} {Ph.D. thesis},\ \bibinfo  {school} {The University
  of Manchester} (\bibinfo {year} {2013})\BibitemShut {NoStop}%
\bibitem [{\citenamefont {De~Witte}\ \emph {et~al.}(2005)\citenamefont
  {De~Witte} \emph {et~al.}}]{DeWitte2005}%
  \BibitemOpen
  \bibfield  {author} {\bibinfo {author} {\bibfnamefont {H.}~\bibnamefont
  {De~Witte}} \emph {et~al.},\ }\href@noop {} {\bibfield  {journal} {\bibinfo
  {journal} {Eur. Phys. J. A}\ }\textbf {\bibinfo {volume} {23}},\ \bibinfo
  {pages} {243} (\bibinfo {year} {2005})}\BibitemShut {NoStop}%
\bibitem [{\citenamefont {Zhu}\ and\ \citenamefont {Kondev}(2008)}]{Zhu2008}%
  \BibitemOpen
  \bibfield  {author} {\bibinfo {author} {\bibfnamefont {S.}~\bibnamefont
  {Zhu}}\ and\ \bibinfo {author} {\bibfnamefont {F.~G.}\ \bibnamefont
  {Kondev}},\ }\href@noop {} {\bibfield  {journal} {\bibinfo  {journal}
  {Nuclear Data Sheets}\ }\textbf {\bibinfo {volume} {109}},\ \bibinfo {pages}
  {699 } (\bibinfo {year} {2008})}\BibitemShut {NoStop}%
\bibitem [{\citenamefont {Lourens}(1967)}]{LourensThesis}%
  \BibitemOpen
  \bibfield  {author} {\bibinfo {author} {\bibfnamefont {W.}~\bibnamefont
  {Lourens}},\ }\href@noop {} {Ph.D. thesis},\ \bibinfo  {school} {Technische
  Hogeschool Delft} (\bibinfo {year} {1967})\BibitemShut {NoStop}%
\bibitem [{\citenamefont {Ritchie}\ \emph {et~al.}(1981)\citenamefont
  {Ritchie}, \citenamefont {Toth}, \citenamefont {Carter}, \citenamefont
  {Mlekodaj},\ and\ \citenamefont {Spejewski}}]{Ritchie1981}%
  \BibitemOpen
  \bibfield  {author} {\bibinfo {author} {\bibfnamefont {B.~G.}\ \bibnamefont
  {Ritchie}}, \bibinfo {author} {\bibfnamefont {K.~S.}\ \bibnamefont {Toth}},
  \bibinfo {author} {\bibfnamefont {H.~K.}\ \bibnamefont {Carter}}, \bibinfo
  {author} {\bibfnamefont {R.~L.}\ \bibnamefont {Mlekodaj}}, \ and\ \bibinfo
  {author} {\bibfnamefont {E.~H.}\ \bibnamefont {Spejewski}},\ }\href@noop {}
  {\bibfield  {journal} {\bibinfo  {journal} {Phys. Rev. C}\ }\textbf {\bibinfo
  {volume} {23}},\ \bibinfo {pages} {2342} (\bibinfo {year}
  {1981})}\BibitemShut {NoStop}%
\bibitem [{\citenamefont {Kondev}\ and\ \citenamefont
  {Lalkovski}(2011)}]{Kondev2011}%
  \BibitemOpen
  \bibfield  {author} {\bibinfo {author} {\bibfnamefont {F.}~\bibnamefont
  {Kondev}}\ and\ \bibinfo {author} {\bibfnamefont {S.}~\bibnamefont
  {Lalkovski}},\ }\href {\doibase http://dx.doi.org/10.1016/j.nds.2011.02.002}
  {\bibfield  {journal} {\bibinfo  {journal} {Nuclear Data Sheets}\ }\textbf
  {\bibinfo {volume} {112}},\ \bibinfo {pages} {707 } (\bibinfo {year}
  {2011})}\BibitemShut {NoStop}%
\bibitem [{\citenamefont {Singh}\ \emph {et~al.}(2013)\citenamefont {Singh},
  \citenamefont {Abriola}, \citenamefont {Baglin}, \citenamefont {Demetriou},
  \citenamefont {Johnson}, \citenamefont {McCutchan}, \citenamefont
  {Mukherjee}, \citenamefont {Singh}, \citenamefont {Sonzogni},\ and\
  \citenamefont {Tuli}}]{Singh2013}%
  \BibitemOpen
  \bibfield  {author} {\bibinfo {author} {\bibfnamefont {B.}~\bibnamefont
  {Singh}}, \bibinfo {author} {\bibfnamefont {D.}~\bibnamefont {Abriola}},
  \bibinfo {author} {\bibfnamefont {C.}~\bibnamefont {Baglin}}, \bibinfo
  {author} {\bibfnamefont {V.}~\bibnamefont {Demetriou}}, \bibinfo {author}
  {\bibfnamefont {T.}~\bibnamefont {Johnson}}, \bibinfo {author} {\bibfnamefont
  {E.}~\bibnamefont {McCutchan}}, \bibinfo {author} {\bibfnamefont
  {G.}~\bibnamefont {Mukherjee}}, \bibinfo {author} {\bibfnamefont
  {S.}~\bibnamefont {Singh}}, \bibinfo {author} {\bibfnamefont
  {A.}~\bibnamefont {Sonzogni}}, \ and\ \bibinfo {author} {\bibfnamefont
  {J.}~\bibnamefont {Tuli}},\ }\href {\doibase
  http://dx.doi.org/10.1016/j.nds.2013.05.001} {\bibfield  {journal} {\bibinfo
  {journal} {Nuclear Data Sheets}\ }\textbf {\bibinfo {volume} {114}},\
  \bibinfo {pages} {661 } (\bibinfo {year} {2013})}\BibitemShut {NoStop}%
\bibitem [{\citenamefont {Browne}\ and\ \citenamefont
  {Tuli}(2011)}]{Browne2011}%
  \BibitemOpen
  \bibfield  {author} {\bibinfo {author} {\bibfnamefont {E.}~\bibnamefont
  {Browne}}\ and\ \bibinfo {author} {\bibfnamefont {J.}~\bibnamefont {Tuli}},\
  }\href {\doibase http://dx.doi.org/10.1016/j.nds.2011.03.002} {\bibfield
  {journal} {\bibinfo  {journal} {Nuclear Data Sheets}\ }\textbf {\bibinfo
  {volume} {112}},\ \bibinfo {pages} {1115 } (\bibinfo {year}
  {2011})}\BibitemShut {NoStop}%
\bibitem [{\citenamefont {Gomez}\ \emph {et~al.}(2008)\citenamefont {Gomez},
  \citenamefont {Aubin}, \citenamefont {Orozco}, \citenamefont {Sprouse},
  \citenamefont {Iskrenova-Tchoukova},\ and\ \citenamefont
  {Safronova}}]{Gomez2008}%
  \BibitemOpen
  \bibfield  {author} {\bibinfo {author} {\bibfnamefont {E.}~\bibnamefont
  {Gomez}}, \bibinfo {author} {\bibfnamefont {S.}~\bibnamefont {Aubin}},
  \bibinfo {author} {\bibfnamefont {L.~A.}\ \bibnamefont {Orozco}}, \bibinfo
  {author} {\bibfnamefont {G.~D.}\ \bibnamefont {Sprouse}}, \bibinfo {author}
  {\bibfnamefont {E.}~\bibnamefont {Iskrenova-Tchoukova}}, \ and\ \bibinfo
  {author} {\bibfnamefont {M.~S.}\ \bibnamefont {Safronova}},\ }\href@noop {}
  {\bibfield  {journal} {\bibinfo  {journal} {Phys. Rev. Lett.}\ }\textbf
  {\bibinfo {volume} {100}},\ \bibinfo {pages} {172502} (\bibinfo {year}
  {2008})}\BibitemShut {NoStop}%
\bibitem [{\citenamefont {Dzuba}\ \emph {et~al.}(2005)\citenamefont {Dzuba},
  \citenamefont {Johnson},\ and\ \citenamefont {Safronova}}]{Dzuba2005}%
  \BibitemOpen
  \bibfield  {author} {\bibinfo {author} {\bibfnamefont {V.~A.}\ \bibnamefont
  {Dzuba}}, \bibinfo {author} {\bibfnamefont {W.~R.}\ \bibnamefont {Johnson}},
  \ and\ \bibinfo {author} {\bibfnamefont {M.~S.}\ \bibnamefont {Safronova}},\
  }\href@noop {} {\bibfield  {journal} {\bibinfo  {journal} {Phys. Rev. A}\
  }\textbf {\bibinfo {volume} {72}},\ \bibinfo {pages} {022503} (\bibinfo
  {year} {2005})}\BibitemShut {NoStop}%
\bibitem [{\citenamefont {King}(1963)}]{King1963}%
  \BibitemOpen
  \bibfield  {author} {\bibinfo {author} {\bibfnamefont {W.~H.}\ \bibnamefont
  {King}},\ }\href@noop {} {\bibfield  {journal} {\bibinfo  {journal} {J. Opt.
  Soc. Am.}\ }\textbf {\bibinfo {volume} {53}},\ \bibinfo {pages} {638}
  (\bibinfo {year} {1963})}\BibitemShut {NoStop}%
\bibitem [{\citenamefont {Lynch}\ \emph
  {et~al.}(2013{\natexlab{b}})\citenamefont {Lynch} \emph
  {et~al.}}]{Lynch2013b}%
  \BibitemOpen
  \bibfield  {author} {\bibinfo {author} {\bibfnamefont {K.~M.}\ \bibnamefont
  {Lynch}} \emph {et~al.},\ }\href {\doibase 10.1051/epjconf/20136301007}
  {\bibfield  {journal} {\bibinfo  {journal} {EPJ Web of Conferences}\ }\textbf
  {\bibinfo {volume} {63}},\ \bibinfo {pages} {01007} (\bibinfo {year}
  {2013}{\natexlab{b}})}\BibitemShut {NoStop}%
\bibitem [{\citenamefont {Ekstr\"om}\ \emph {et~al.}(1986)\citenamefont
  {Ekstr\"om}, \citenamefont {Robertsson},\ and\ \citenamefont
  {Ros\'en}}]{Ekstrom1986}%
  \BibitemOpen
  \bibfield  {author} {\bibinfo {author} {\bibfnamefont {C.}~\bibnamefont
  {Ekstr\"om}}, \bibinfo {author} {\bibfnamefont {L.}~\bibnamefont
  {Robertsson}}, \ and\ \bibinfo {author} {\bibfnamefont {A.}~\bibnamefont
  {Ros\'en}},\ }\href@noop {} {\bibfield  {journal} {\bibinfo  {journal} {Phys.
  Scripta}\ }\textbf {\bibinfo {volume} {34}},\ \bibinfo {pages} {624}
  (\bibinfo {year} {1986})}\BibitemShut {NoStop}%
\bibitem [{\citenamefont {Stroke}\ \emph {et~al.}(1961)\citenamefont {Stroke},
  \citenamefont {Blin-Stoyle},\ and\ \citenamefont {Jaccarino}}]{Stroke1961}%
  \BibitemOpen
  \bibfield  {author} {\bibinfo {author} {\bibfnamefont {H.~H.}\ \bibnamefont
  {Stroke}}, \bibinfo {author} {\bibfnamefont {R.~J.}\ \bibnamefont
  {Blin-Stoyle}}, \ and\ \bibinfo {author} {\bibfnamefont {V.}~\bibnamefont
  {Jaccarino}},\ }\href@noop {} {\bibfield  {journal} {\bibinfo  {journal}
  {Phys. Rev.}\ }\textbf {\bibinfo {volume} {123}},\ \bibinfo {pages} {1326}
  (\bibinfo {year} {1961})}\BibitemShut {NoStop}%
\bibitem [{\citenamefont {Anselment}\ \emph {et~al.}(1986)\citenamefont
  {Anselment}, \citenamefont {Faubel}, \citenamefont {G\"oring}, \citenamefont
  {Hanser}, \citenamefont {Meisel}, \citenamefont {Rebel},\ and\ \citenamefont
  {Schatz}}]{Anselment1986}%
  \BibitemOpen
  \bibfield  {author} {\bibinfo {author} {\bibfnamefont {M.}~\bibnamefont
  {Anselment}}, \bibinfo {author} {\bibfnamefont {W.}~\bibnamefont {Faubel}},
  \bibinfo {author} {\bibfnamefont {S.}~\bibnamefont {G\"oring}}, \bibinfo
  {author} {\bibfnamefont {A.}~\bibnamefont {Hanser}}, \bibinfo {author}
  {\bibfnamefont {G.}~\bibnamefont {Meisel}}, \bibinfo {author} {\bibfnamefont
  {H.}~\bibnamefont {Rebel}}, \ and\ \bibinfo {author} {\bibfnamefont
  {G.}~\bibnamefont {Schatz}},\ }\href@noop {} {\bibfield  {journal} {\bibinfo
  {journal} {Nucl. Phys. A}\ }\textbf {\bibinfo {volume} {451}},\ \bibinfo
  {pages} {471 } (\bibinfo {year} {1986})}\BibitemShut {NoStop}%
\bibitem [{\citenamefont {Myers}\ and\ \citenamefont
  {Schmidt}(1983)}]{Myers1983}%
  \BibitemOpen
  \bibfield  {author} {\bibinfo {author} {\bibfnamefont {W.~D.}\ \bibnamefont
  {Myers}}\ and\ \bibinfo {author} {\bibfnamefont {K.-H.}\ \bibnamefont
  {Schmidt}},\ }\href@noop {} {\bibfield  {journal} {\bibinfo  {journal} {Nucl.
  Phys. A}\ }\textbf {\bibinfo {volume} {410}},\ \bibinfo {pages} {61 }
  (\bibinfo {year} {1983})}\BibitemShut {NoStop}%
\bibitem [{\citenamefont {Raman}\ \emph {et~al.}(2001)\citenamefont {Raman},
  \citenamefont {Jr.},\ and\ \citenamefont {Tikkanen}}]{Raman2001}%
  \BibitemOpen
  \bibfield  {author} {\bibinfo {author} {\bibfnamefont {S.}~\bibnamefont
  {Raman}}, \bibinfo {author} {\bibfnamefont {C.~N.}\ \bibnamefont {Jr.}}, \
  and\ \bibinfo {author} {\bibfnamefont {P.}~\bibnamefont {Tikkanen}},\ }\href
  {\doibase http://dx.doi.org/10.1006/adnd.2001.0858} {\bibfield  {journal}
  {\bibinfo  {journal} {Atomic Data and Nuclear Data Tables}\ }\textbf
  {\bibinfo {volume} {78}},\ \bibinfo {pages} {1 } (\bibinfo {year}
  {2001})}\BibitemShut {NoStop}%
\bibitem [{\citenamefont {De~Witte}\ \emph {et~al.}(2007)\citenamefont
  {De~Witte} \emph {et~al.}}]{DeWitte2007}%
  \BibitemOpen
  \bibfield  {author} {\bibinfo {author} {\bibfnamefont {H.}~\bibnamefont
  {De~Witte}} \emph {et~al.},\ }\href@noop {} {\bibfield  {journal} {\bibinfo
  {journal} {Phys. Rev. Lett.}\ }\textbf {\bibinfo {volume} {98}},\ \bibinfo
  {pages} {112502} (\bibinfo {year} {2007})}\BibitemShut {NoStop}%
\bibitem [{\citenamefont {Borchers}\ \emph {et~al.}(1987)\citenamefont
  {Borchers}, \citenamefont {Neugart}, \citenamefont {Otten}, \citenamefont
  {Duong}, \citenamefont {Ulm},\ and\ \citenamefont {Wendt}}]{Borchers1987}%
  \BibitemOpen
  \bibfield  {author} {\bibinfo {author} {\bibfnamefont {W.}~\bibnamefont
  {Borchers}}, \bibinfo {author} {\bibfnamefont {R.}~\bibnamefont {Neugart}},
  \bibinfo {author} {\bibfnamefont {E.~W.}\ \bibnamefont {Otten}}, \bibinfo
  {author} {\bibfnamefont {H.~T.}\ \bibnamefont {Duong}}, \bibinfo {author}
  {\bibfnamefont {G.}~\bibnamefont {Ulm}}, \ and\ \bibinfo {author}
  {\bibfnamefont {K.}~\bibnamefont {Wendt}},\ }\href@noop {} {\bibfield
  {journal} {\bibinfo  {journal} {Hyperfine Interact.}\ }\textbf {\bibinfo
  {volume} {34}},\ \bibinfo {pages} {25} (\bibinfo {year} {1987})}\BibitemShut
  {NoStop}%
\bibitem [{\citenamefont {Bastug}\ \emph {et~al.}(1996)\citenamefont {Bastug},
  \citenamefont {Fricke}, \citenamefont {Finkbeiner},\ and\ \citenamefont
  {Johnson}}]{Bastug1996}%
  \BibitemOpen
  \bibfield  {author} {\bibinfo {author} {\bibfnamefont {T.}~\bibnamefont
  {Bastug}}, \bibinfo {author} {\bibfnamefont {B.}~\bibnamefont {Fricke}},
  \bibinfo {author} {\bibfnamefont {M.}~\bibnamefont {Finkbeiner}}, \ and\
  \bibinfo {author} {\bibfnamefont {W.~R.}\ \bibnamefont {Johnson}},\
  }\href@noop {} {\bibfield  {journal} {\bibinfo  {journal} {Z. Phys. D}\
  }\textbf {\bibinfo {volume} {37}},\ \bibinfo {pages} {281} (\bibinfo {year}
  {1996})}\BibitemShut {NoStop}%
\bibitem [{\citenamefont {Neugart}\ \emph {et~al.}(1985)\citenamefont
  {Neugart}, \citenamefont {Stroke}, \citenamefont {Ahmad}, \citenamefont
  {Duong}, \citenamefont {Ravn},\ and\ \citenamefont {Wendt}}]{Neugart1985}%
  \BibitemOpen
  \bibfield  {author} {\bibinfo {author} {\bibfnamefont {R.}~\bibnamefont
  {Neugart}}, \bibinfo {author} {\bibfnamefont {H.~H.}\ \bibnamefont {Stroke}},
  \bibinfo {author} {\bibfnamefont {S.~A.}\ \bibnamefont {Ahmad}}, \bibinfo
  {author} {\bibfnamefont {H.~T.}\ \bibnamefont {Duong}}, \bibinfo {author}
  {\bibfnamefont {H.~L.}\ \bibnamefont {Ravn}}, \ and\ \bibinfo {author}
  {\bibfnamefont {K.}~\bibnamefont {Wendt}},\ }\href {\doibase
  10.1103/PhysRevLett.55.1559} {\bibfield  {journal} {\bibinfo  {journal}
  {Phys. Rev. Lett.}\ }\textbf {\bibinfo {volume} {55}},\ \bibinfo {pages}
  {1559} (\bibinfo {year} {1985})}\BibitemShut {NoStop}%
\bibitem [{\citenamefont {Stone}(2011)}]{Stone2011}%
  \BibitemOpen
  \bibfield  {author} {\bibinfo {author} {\bibfnamefont {N.}~\bibnamefont
  {Stone}},\ }\href@noop {} {\emph {\bibinfo {title} {Table of Nuclear Magnetic
  Dipole and Electric Quadrupole Moments}}}\ (\bibinfo  {publisher} {Nuclear
  Data Services},\ \bibinfo {address} {International Atomic Energy Agency,
  Vienna, Austria},\ \bibinfo {year} {2011})\BibitemShut {NoStop}%
\bibitem [{\citenamefont {Barzakh}\ \emph {et~al.}(2012)\citenamefont
  {Barzakh}, \citenamefont {Batist}, \citenamefont {Fedorov}, \citenamefont
  {Ivanov}, \citenamefont {Mezilev}, \citenamefont {Molkanov}, \citenamefont
  {Moroz}, \citenamefont {Orlov}, \citenamefont {Panteleev},\ and\
  \citenamefont {Volkov}}]{Barzakh2012}%
  \BibitemOpen
  \bibfield  {author} {\bibinfo {author} {\bibfnamefont {A.~E.}\ \bibnamefont
  {Barzakh}}, \bibinfo {author} {\bibfnamefont {L.~K.}\ \bibnamefont {Batist}},
  \bibinfo {author} {\bibfnamefont {D.~V.}\ \bibnamefont {Fedorov}}, \bibinfo
  {author} {\bibfnamefont {V.~S.}\ \bibnamefont {Ivanov}}, \bibinfo {author}
  {\bibfnamefont {K.~A.}\ \bibnamefont {Mezilev}}, \bibinfo {author}
  {\bibfnamefont {P.~L.}\ \bibnamefont {Molkanov}}, \bibinfo {author}
  {\bibfnamefont {F.~V.}\ \bibnamefont {Moroz}}, \bibinfo {author}
  {\bibfnamefont {S.~Y.}\ \bibnamefont {Orlov}}, \bibinfo {author}
  {\bibfnamefont {V.~N.}\ \bibnamefont {Panteleev}}, \ and\ \bibinfo {author}
  {\bibfnamefont {Y.~M.}\ \bibnamefont {Volkov}},\ }\href {\doibase
  10.1103/PhysRevC.86.014311} {\bibfield  {journal} {\bibinfo  {journal} {Phys.
  Rev. C}\ }\textbf {\bibinfo {volume} {86}},\ \bibinfo {pages} {014311}
  (\bibinfo {year} {2012})}\BibitemShut {NoStop}%
\bibitem [{\citenamefont {Neyens}(2003)}]{Neyens2003}%
  \BibitemOpen
  \bibfield  {author} {\bibinfo {author} {\bibfnamefont {G.}~\bibnamefont
  {Neyens}},\ }\href@noop {} {\bibfield  {journal} {\bibinfo  {journal} {Rep.
  Prog. Phys.}\ }\textbf {\bibinfo {volume} {66}},\ \bibinfo {pages} {633}
  (\bibinfo {year} {2003})}\BibitemShut {NoStop}%
\bibitem [{\citenamefont {Wouters}\ \emph {et~al.}(1991)\citenamefont
  {Wouters}, \citenamefont {Severijns}, \citenamefont {Vanhaverbeke},\ and\
  \citenamefont {Vanneste}}]{Wouters1991}%
  \BibitemOpen
  \bibfield  {author} {\bibinfo {author} {\bibfnamefont {J.}~\bibnamefont
  {Wouters}}, \bibinfo {author} {\bibfnamefont {N.}~\bibnamefont {Severijns}},
  \bibinfo {author} {\bibfnamefont {J.}~\bibnamefont {Vanhaverbeke}}, \ and\
  \bibinfo {author} {\bibfnamefont {L.}~\bibnamefont {Vanneste}},\ }\href@noop
  {} {\bibfield  {journal} {\bibinfo  {journal} {J. Phys. G: Nucl. Part.
  Phys.}\ }\textbf {\bibinfo {volume} {17}},\ \bibinfo {pages} {1673} (\bibinfo
  {year} {1991})}\BibitemShut {NoStop}%
\bibitem [{\citenamefont {Ahmad}\ \emph {et~al.}(1983)\citenamefont {Ahmad},
  \citenamefont {Klempt}, \citenamefont {Neugart}, \citenamefont {Otten},
  \citenamefont {Wendt},\ and\ \citenamefont {Ekstr\"om}}]{Ahmad1983}%
  \BibitemOpen
  \bibfield  {author} {\bibinfo {author} {\bibfnamefont {S.~A.}\ \bibnamefont
  {Ahmad}}, \bibinfo {author} {\bibfnamefont {W.}~\bibnamefont {Klempt}},
  \bibinfo {author} {\bibfnamefont {R.}~\bibnamefont {Neugart}}, \bibinfo
  {author} {\bibfnamefont {E.~W.}\ \bibnamefont {Otten}}, \bibinfo {author}
  {\bibfnamefont {K.}~\bibnamefont {Wendt}}, \ and\ \bibinfo {author}
  {\bibfnamefont {C.}~\bibnamefont {Ekstr\"om}},\ }\href@noop {} {\bibfield
  {journal} {\bibinfo  {journal} {Physics Letters B}\ }\textbf {\bibinfo
  {volume} {133}},\ \bibinfo {pages} {47 } (\bibinfo {year}
  {1983})}\BibitemShut {NoStop}%
\bibitem [{\citenamefont {Duppen}\ \emph {et~al.}(1991)\citenamefont {Duppen},
  \citenamefont {Decrock}, \citenamefont {Dendooven}, \citenamefont {Huyse},
  \citenamefont {Reusen},\ and\ \citenamefont {Wauters}}]{VanDuppen1991}%
  \BibitemOpen
  \bibfield  {author} {\bibinfo {author} {\bibfnamefont {P.~V.}\ \bibnamefont
  {Duppen}}, \bibinfo {author} {\bibfnamefont {P.}~\bibnamefont {Decrock}},
  \bibinfo {author} {\bibfnamefont {P.}~\bibnamefont {Dendooven}}, \bibinfo
  {author} {\bibfnamefont {M.}~\bibnamefont {Huyse}}, \bibinfo {author}
  {\bibfnamefont {G.}~\bibnamefont {Reusen}}, \ and\ \bibinfo {author}
  {\bibfnamefont {J.}~\bibnamefont {Wauters}},\ }\href@noop {} {\bibfield
  {journal} {\bibinfo  {journal} {Nucl. Phys. A}\ }\textbf {\bibinfo {volume}
  {529}},\ \bibinfo {pages} {268 } (\bibinfo {year} {1991})}\BibitemShut
  {NoStop}%
\bibitem [{\citenamefont {Kondev}(2008)}]{Kondev2008}%
  \BibitemOpen
  \bibfield  {author} {\bibinfo {author} {\bibfnamefont {F.~G.}\ \bibnamefont
  {Kondev}},\ }\href@noop {} {\bibfield  {journal} {\bibinfo  {journal}
  {Nuclear Data Sheets}\ }\textbf {\bibinfo {volume} {109}},\ \bibinfo {pages}
  {1527 } (\bibinfo {year} {2008})}\BibitemShut {NoStop}%
\bibitem [{\citenamefont {Chiara}\ and\ \citenamefont
  {Kondev}(2010)}]{Chiara2010}%
  \BibitemOpen
  \bibfield  {author} {\bibinfo {author} {\bibfnamefont {C.~J.}\ \bibnamefont
  {Chiara}}\ and\ \bibinfo {author} {\bibfnamefont {F.~G.}\ \bibnamefont
  {Kondev}},\ }\href@noop {} {\bibfield  {journal} {\bibinfo  {journal}
  {Nuclear Data Sheets}\ }\textbf {\bibinfo {volume} {111}},\ \bibinfo {pages}
  {141 } (\bibinfo {year} {2010})}\BibitemShut {NoStop}%
\bibitem [{\citenamefont {Martin}(2007)}]{Martin2007}%
  \BibitemOpen
  \bibfield  {author} {\bibinfo {author} {\bibfnamefont {M.~J.}\ \bibnamefont
  {Martin}},\ }\href@noop {} {\bibfield  {journal} {\bibinfo  {journal}
  {Nuclear Data Sheets}\ }\textbf {\bibinfo {volume} {108}},\ \bibinfo {pages}
  {1583 } (\bibinfo {year} {2007})}\BibitemShut {NoStop}%
\bibitem [{\citenamefont {Browne}(2005)}]{Browne2005}%
  \BibitemOpen
  \bibfield  {author} {\bibinfo {author} {\bibfnamefont {E.}~\bibnamefont
  {Browne}},\ }\href@noop {} {\bibfield  {journal} {\bibinfo  {journal}
  {Nuclear Data Sheets}\ }\textbf {\bibinfo {volume} {104}},\ \bibinfo {pages}
  {427 } (\bibinfo {year} {2005})}\BibitemShut {NoStop}%
\bibitem [{\citenamefont {Lynch}\ \emph {et~al.}(2012)\citenamefont {Lynch}
  \emph {et~al.}}]{Lynch2012}%
  \BibitemOpen
  \bibfield  {author} {\bibinfo {author} {\bibfnamefont {K.~M.}\ \bibnamefont
  {Lynch}} \emph {et~al.},\ }\href@noop {} {\bibfield  {journal} {\bibinfo
  {journal} {J. Phys.: Conf. Ser.}\ }\textbf {\bibinfo {volume} {381}},\
  \bibinfo {pages} {012128} (\bibinfo {year} {2012})}\BibitemShut {NoStop}%
\end{thebibliography}%

\end{document}